\documentclass[english,pra,superscriptaddress,reprint,twocolumn]{revtex4-1}
\usepackage{standalone}
\usepackage{nccfoots}
\usepackage{comment}
\usepackage{graphics}
\usepackage{float}
\graphicspath{{figures/}}
\usepackage{braket}
\usepackage{color}
\usepackage{nicefrac}
\usepackage{mathtools}
\usepackage{bbold}

\usepackage[colorlinks=true,
   linkcolor=blue,
   citecolor=blue,
   urlcolor =blue]{hyperref}

\newcommand{\citeSM}{\cite[{\tiny SM}\kern-0.3em][]{SM}}

\newcommand{\be}{\begin{equation}}
\newcommand{\ee}{\end{equation}}

\newcommand{\tr}{\mathrm{Tr}}
\newcommand{\var}{\mathrm{Var}}
\newcommand{\cov}{\mathrm{Cov}}
\newcommand{\expp}{\mathbb{E}}
\newcommand{\ar}[1]{\textcolor{black}{#1}}
\newcommand{\aar}[1]{\textcolor{black}{#1}}

\expandafter\let\csname equation*\endcsname\relax

\expandafter\let\csname endequation*\endcsname\relax

\usepackage{amsmath,amsfonts,amssymb,amsthm, bbm,braket}
\usepackage{tikz}
\usepackage{xcolor}

\usepackage[normalem]{ulem}

\DeclarePairedDelimiter\ceil{\lceil}{\rceil}
\DeclarePairedDelimiter\floor{\lfloor}{\rfloor}

\definecolor{Red}{HTML}{E53E30}  
\definecolor{Green}{HTML}{00AD69}  
\definecolor{Blue}{HTML}{2171b5}
\definecolor{Purple}{HTML}{652F6C}  

\newcommand{\bv}[1]{\textcolor{black}{#1}}
\newcommand{\bvv}[1]{\textcolor{black}{#1}}
\newcommand{\cb}[1]{\textcolor{black}{#1}}
\newcommand{\am}[1]{\textcolor{black}{#1}}

\begin{document}
\title{Quantum Fisher information from randomized measurements}
\author{Aniket Rath}
\affiliation{Universit\'e  Grenoble Alpes, CNRS, LPMMC, 38000 Grenoble, France}

\author{Cyril Branciard}
\affiliation{Universit\'e Grenoble Alpes, CNRS, Grenoble INP, Institut N\'eel, 38000 Grenoble, France}

\author{Anna Minguzzi}
\affiliation{Universit\'e  Grenoble Alpes, CNRS, LPMMC, 38000 Grenoble, France}

\author{Beno\^it Vermersch}
\affiliation{Universit\'e  Grenoble Alpes, CNRS, LPMMC, 38000 Grenoble, France}
\affiliation{Center for Quantum Physics, University of Innsbruck, Innsbruck A-6020, Austria}	
\affiliation{Institute for Quantum Optics and Quantum Information of the Austrian Academy of Sciences,  Innsbruck A-6020, Austria}

\begin{abstract}
The quantum Fisher information (QFI) is a fundamental quantity of interest in many areas
from quantum metrology to quantum information theory. It can in particular be used as a  witness to establish the degree of multi-particle entanglement in quantum many-body-systems. In this work, we use polynomials of the density matrix to construct monotonically increasing lower bounds that converge to the QFI. Using randomized measurements we propose a protocol to accurately estimate these lower bounds in state-of-art quantum technological platforms. We 
\am{estimate the number of measurements needed to achieve a given accuracy and confidence level in the bounds, and}
present two examples of applications of the method in quantum systems made of coupled qubits and collective spins.

\end{abstract}
\maketitle

First introduced in quantum metrology to measure the ability for quantum states to perform interferometry beyond the shot-noise limit~\cite{Caves1994,Braunstein1996}, the quantum Fisher information (QFI) \bv{plays a fundamental role} in different fields, including quantum information theory and many-body physics.
\aar{As enhanced sensitivity for metrology and sensing requires the generation of multipartite entangled states~\cite{Pezze2009}}, the QFI has raised significant interest as a witness of entanglement. In particular, the notion of entanglement `depth'---the minimum number of entangled particles in a given state---\ar{and the underlying structure of multipartite entanglement }can be related to the value of the QFI~\cite{Pezze2018,Ren2021}. In many-body physics, the ability for the QFI to reveal the entanglement of mixed states makes it a key quantity in the study of spin models, revealing in particular universal entanglement properties of quantum states crossing a phase transition at finite temperature~\cite{Zanardi2008} \ar{and \cb{highlighting} the role of multipartite entanglement in topological phase transitions~\cite{Pezze2017}}. 
This letter presents a protocol to estimate the QFI in state-of-art quantum devices via randomized measurements. 

The challenge to measure the QFI arises from it being a highly non-linear function of the density matrix. 
\ar{The QFI is defined with respect to a given Hermitian operator $A$ and a quantum state $\rho$, and can be written in the following closed form}
\begin{equation}
    F_Q = 2 \,\tr \bigg( \frac{(\rho \otimes \mathbb{1} - \mathbb{1} \otimes \rho)^2}{\rho \otimes \mathbb{1} + \mathbb{1} \otimes \rho} \mathbb{S} (A \otimes A) \bigg) \label{QFI},
\end{equation}
where $\mathbb{S}$ is the swap operator defined through its action on basis states $\ket{i_1}$, $\ket{i_2}$ by $\ar{\mathbb{S}(\ket{i_1} \otimes \ket{i_2}) = \ket{i_2} \otimes \ket{i_1}}$.
We clarify the form of Eq.~\eqref{QFI}---noting that the fraction notation is allowed by the fact that the numerator and denominator commute---and relate it to the standard expression of the QFI: \bv{
$ F_Q = 2\sum_{(i,j),\lambda_i + \lambda_j>0} \frac{(\lambda_i - \lambda_j)^2}{\lambda_i + \lambda_j} |\bra{i}A\ket{j}|^2$,
with $\rho=\sum_i \lambda_i \ket{i} \bra{i}$,
in the Supplemental Material (SM)~\cite{SM}\nocite{Caves1994,Braunstein1996,Hoeffding1992,Huang2020,Elben2020b}}. For pure states $\rho=\ket{\psi}\bra{\psi}$, the QFI is proportional to the variance of the operator $A$, \mbox{$F_Q = 4(\bra{\psi}A^2\ket{\psi}-\bra{\psi}A\ket{\psi}^2)$}.
\bv{
According to the 
 quantum Cram\'er-Rao relation, the QFI bounds the achievable precision in parameter estimation in quantum metrology~\cite{Pezze2018}.
Furthermore, for $N$ spins-$1/2$,  with a collective spin operator $A=\frac{1}{2} \sum_{l = 1}^{N} \sigma_{\mu}^{(l)}$~\footnote{$\sigma_{\mu}^{(l)}$ is the Pauli matrix in an arbitrary direction $\mu$ acting on the $l^{th}$ spin}, all separable states satisfy $F_Q\le N$~\cite{Pezze2009}.
This means that, if $F_Q>N$, 
the state is entangled, and provides an advantage over all separable states for performing quantum metrology.
}
\bv{The QFI can also be used to certify 
 multipartite entanglement in terms of  $k-$producibility}, \am{i.e.} a decomposition into a statistical mixture of tensor products of $k$-particle states,
 or $m-$separability, \am{i.e.} a decomposition into a statistical mixture of products of at least $m$ factors, of the state $\rho$~\cite{Toth2012,Hyllus2012,Ren2021}.
In particular, the inequality  $F_Q > \Gamma(N,k)$, with \mbox{$\Gamma(N,k) = \floor*{\frac{N}{k}}k^2 + \big(N - \floor*{\frac{N}{k}}k\big)^2$}, implies that a state is not $k$-producible, i.e that it has an entanglement depth of at least $k+1$.
\ar{Note that} one can show the QFI being above a certain threshold value by measuring a \emph{lower bound} of it. This includes quantities associated with the expectation value of an observable, such as spin squeezing~\cite{Monz2011,Strobel2014,Barontini2015,Bohnet2016,Schmied2016,Pezze2009}, or multiple quantum coherence~\cite{Garttner2018}. Recently, non-linear lower bounds to the QFI, not accessible by standard observable measurements, have also been introduced~\cite{Rivas2008,Rivas2010,Zhang2017,Girolami2017,Beckey2020,Cerezo2021} and measured~\cite{Yu2021}.
However, the finite distance between these bounds and the QFI typically limits the ability to certify quantum states for metrology, or to detect multipartite entanglement. 
If the quantum state is in a thermal state, the QFI can also be measured via dynamical susceptibilities~\cite{Garttner2017}. However, the states used in the context of quantum metrology~\cite{Pezze2018}, and many-body dynamics~\cite{Lewis-Swan2019}, are usually out of equilibrium.

Here, we propose a systematic and state-agnostic way to estimate the QFI by measuring a converging series of 
monotonically increasing approximations $F_n$, \mbox{$F_0 \leq F_1 \leq \dots \leq  F_Q$}, which rapidly tend to $F_Q$ as $n$ increases.
Thus, each $F_n$, being a lower-bound to the QFI, allows the verification of quantum metrological advantage and/or multipartite entanglement of the quantum state $\rho$.
Moreover, each function $F_n$, being a \emph{polynomial} function of the density matrix $\rho$, can be accessed by randomized measurements.
Such protocols only require single qubit random rotations and measurements and have been successfully applied to obtain R\'enyi entropies~\cite{vanEnk2012,Elben2018,Brydges2019,Vitale2021,Satzinger2021}, negativities~\cite{Zhou2020,Elben2020b,Neven2021,Xiao2021}, state overlaps~\cite{Elben2020a} (which lead to the sub-QFI, a lower bound on the QFI measured in Ref.~\cite{Yu2021}), scrambling~\cite{Vermersch2019,Joshi2019} and topological invariants~\cite{Elben2019, Cian2020}.
Note that multipartite entanglement conditions can also be expressed \ar{as} statistical moments of randomized measurements~\cite{Knips2019,Ketterer2019,Ketterer2020a,Ketterer2020,Imai2021}. 

{\it Construction of converging lower bounds---}
\bv{We define the bounds $F_n$ as}
\begin{equation}
    F_n = 2 \, \tr \bigg( \sum_{\ell = 0}^{n} (\rho \otimes \mathbb{1} - \mathbb{1} \otimes \rho)^2 (\mathbb{1} \otimes \mathbb{1} -\rho \otimes \mathbb{1} - \mathbb{1} \otimes \rho)^\ell \mathbb{S}(A \otimes A)\bigg). \label{bounds-rho}
\end{equation}
The construction of the above bounds is detailed in the SM~\cite{SM}, where we show that $\forall n \in \mathbb{N}$, $F_n \leq F_Q$ and \mbox{$F_n \leq F_{n+1}$} with the inequalities saturating for pure and fully mixed states.
For the first two bounds, we obtain
\bv{
\begin{align}
    &&
    F_0 &= 4\,\mathrm{Tr}(\rho \big[\rho,A\big] A) \,, \label{F0} \\  
    &&
    F_1 &= 2\,F_0 - 4\,\tr(\rho^2[\rho,A]A) \,. \label{F1}
\end{align}}
\ar{where [,] is the commutator.}
As shown in the SM~\cite{SM}, the constructed series $F_n$ converges exponentially with $n$ to $F_Q$.
Note that the quantity $F_0$ was shown to be a lower bound of the QFI in Refs.~\cite{Rivas2008,Rivas2010,Zhang2017,Girolami2017}, while the sub-QFI of Refs.~\cite{Cerezo2021,Yu2021} \ar{is a lower bound} to $F_0$. \ar{We remark that $F_0$ was proven to be faithful to the QFI with respect to global extrema~\cite{Cerezo2021}.}

\begin{figure}[t]
\begin{minipage}[b]{0.5\linewidth}
\centering
\includegraphics[width=\textwidth]{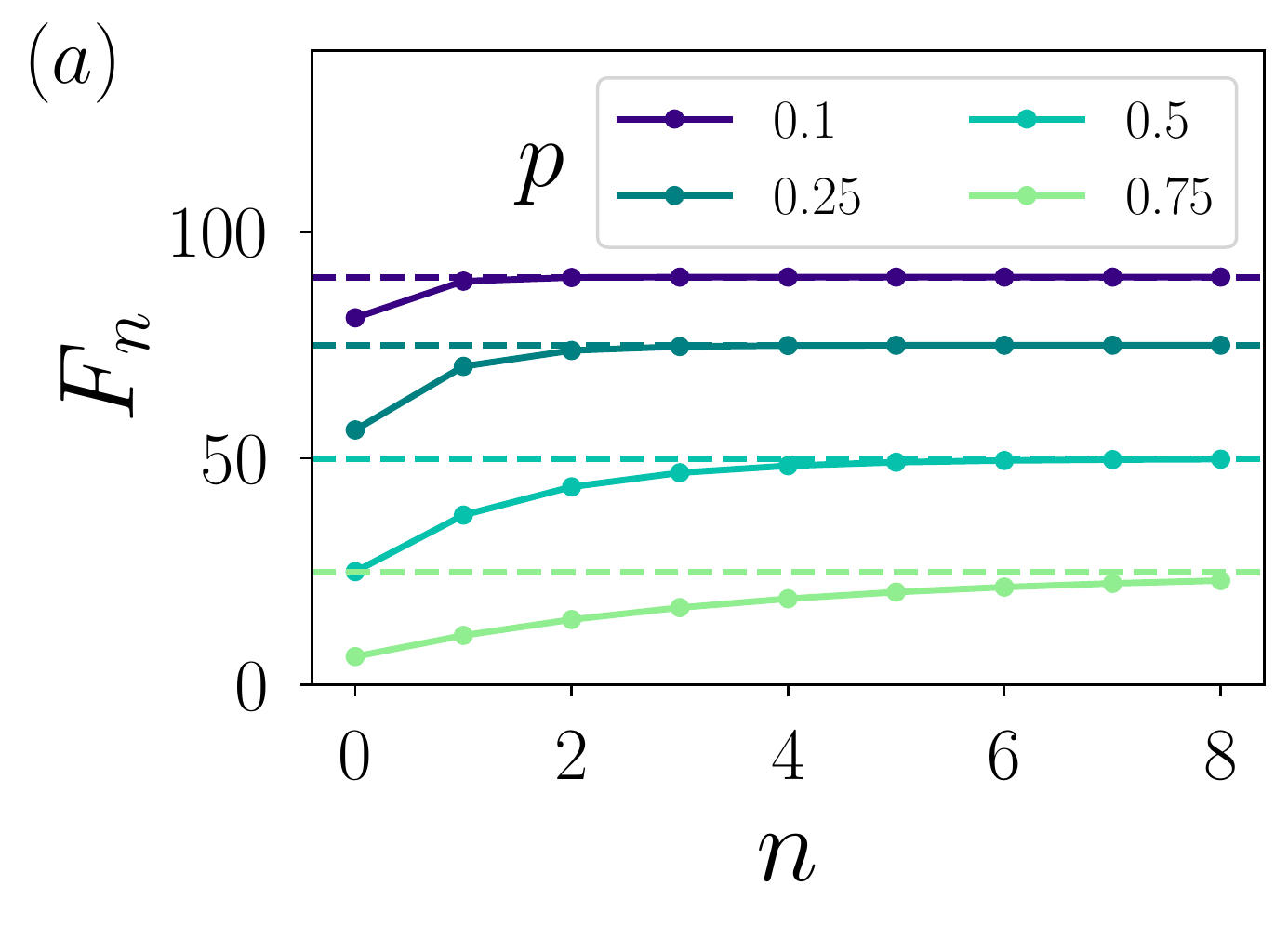}
\end{minipage}
\hskip -1ex
\begin{minipage}[b]{0.5\linewidth}
\centering
\includegraphics[width=\textwidth]{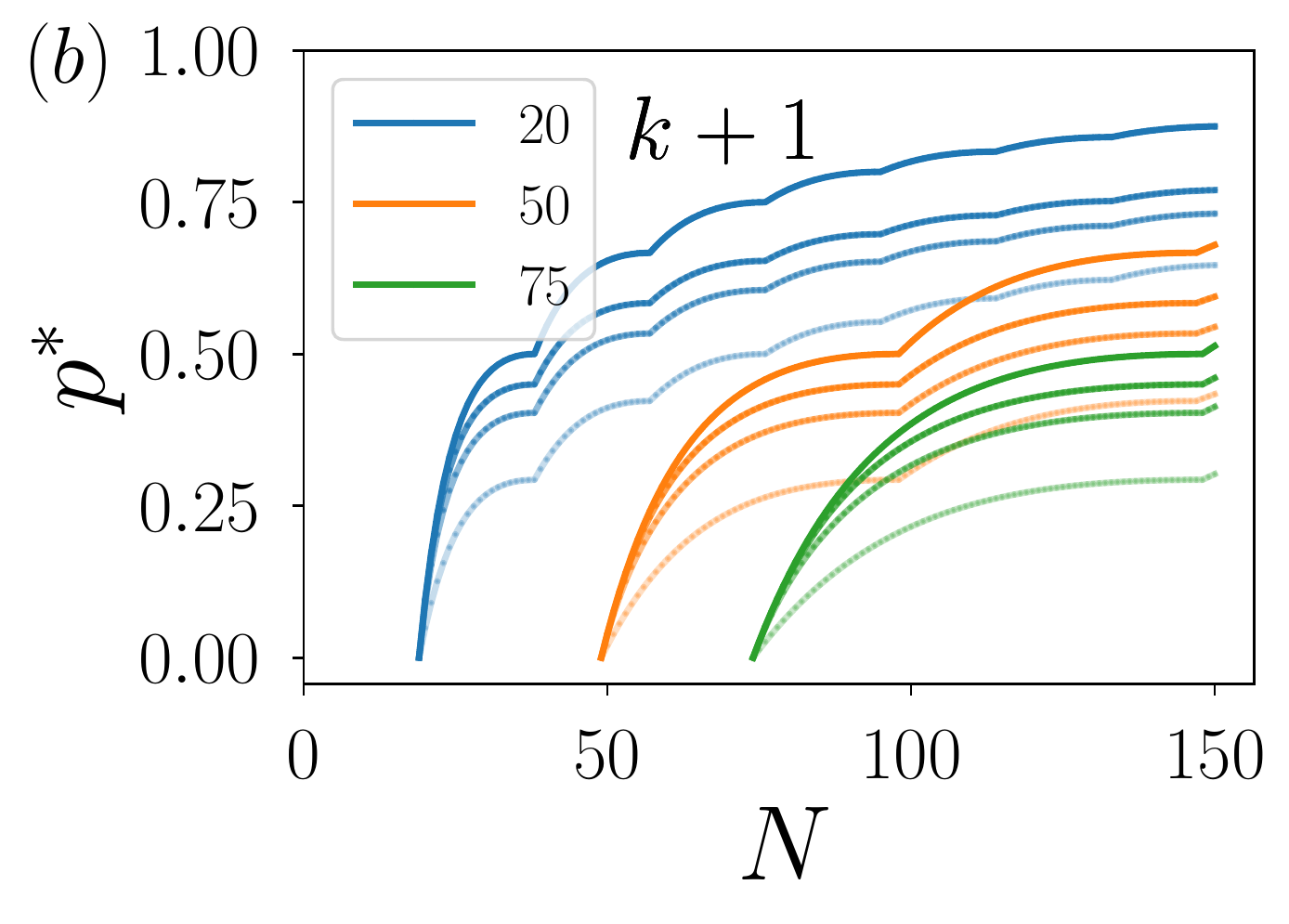}
\end{minipage}
\caption{{\it Convergence of the lower bounds and entanglement depth certification---} 
Panel (a) shows the QFI (dashed lines) and its lower bounds $F_n$ as a function of the order $n$ (dots connected by solid lines) for a 10-qubit GHZ state mixed with different white noise strengths $p$ (see legend). The convergence nature of $F_n$ is commented in the SM~\cite{SM}. Panel (b) shows for noisy GHZ states, the threshold value of white noise $p^*$ as a function of the number of qubits $N$ to detect different entanglement depths of (at least) $k+1$ (see legend) by $F_0$, $F_1$, $F_2$ and $F_Q$ (light to dark).} \label{fig:bounds}
\end{figure}

Fig.~\ref{fig:bounds}(a) illustrates this convergence for different purities of the noisy GHZ state \mbox{$\rho(p) = (1-p)\ket{\mathrm{GHZ}_N}\bra{\mathrm{GHZ}_N} + p\,\mathbb{1}/2^N$} with $\ket{\mathrm{GHZ}_N} = (\ket{0}^{\otimes N}+\ket{1}^{\otimes N})/\sqrt{2}$, and \mbox{$A = \frac{1}{2} \sum_{l = 1}^{N} \sigma_z^{(l)}$} (where to maximize the QFI obtained for this state, we choose the direction $\mu = z$ for $A$). GHZ states of up to 20 qubits have been realized in recent quantum platforms~\cite{Wei2020,Song2019,Omran2019}. This class of states can be used to achieve enhanced sensitivities in quantum metrology~\cite{Pezze2018} as they exhibit non-trivial multipartite entanglement, which cannot be detected using spin-squeezing inequalities~\cite{Pezze2018}. \ar{One of the} important consequences of having a series of monotonically increasing bounds \cb{is that one detects} multipartite entanglement more efficiently as $n$ increases. 
This is illustrated in Fig.~\ref{fig:bounds}(b), where for various values of $N$ and $k$, we consider the maximal value $p^*$, such that an entanglement depth of at least $k+1$ is detected via the inequality $F_n>\Gamma(N,k)$ (which implies  $F_Q>\Gamma(N,k)$).
The noise tolerance $p^*$ increases as a function of the order $n$ of the lower bounds and is upper bounded by the $p^*$ value corresponding to $F_Q$. 

{\it Randomized measurement protocol---}
Let us now show how the bounds $F_n$ can be accessed from randomized measurements. Such protocols first gave access to the purity $\mathrm{Tr}(\rho^2)$~\cite{Elben2018}, and then later to any polynomial of the density matrix~\cite{Huang2020,Elben2020b,Vitale2021,Neven2021}. What makes our bounds $F_n$ accessible from randomized measurements data is precisely that they are polynomials of $\rho$ (of order $n+2$).

\begin{figure}[t]
\centering
\begin{minipage}[b]{0.75\linewidth}
\includegraphics[width=\textwidth]{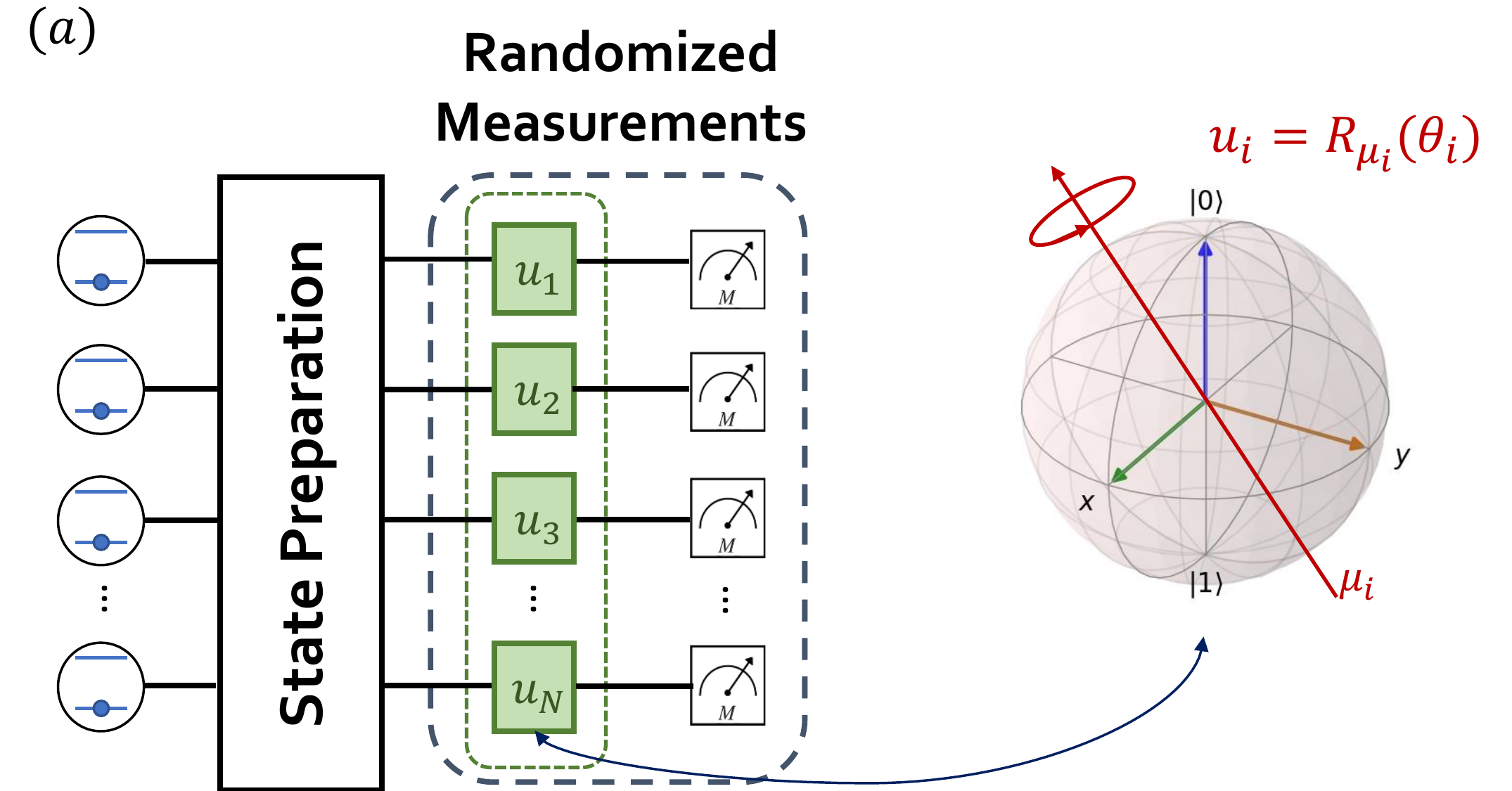}
\end{minipage}
\hskip -1ex
\begin{minipage}[b]{0.5\linewidth}
\centering
\includegraphics[width=\textwidth]{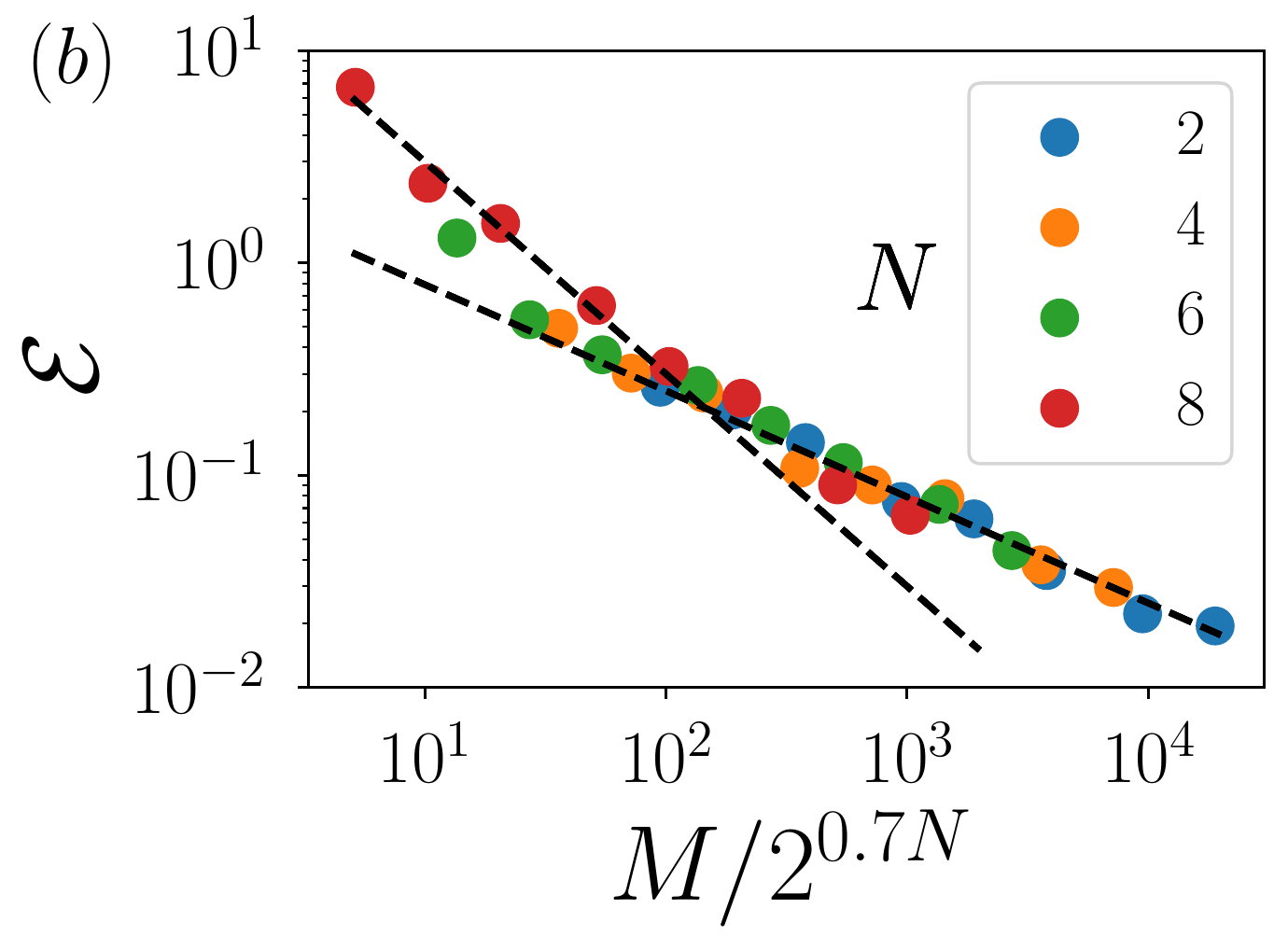}
\end{minipage}
\hskip -1ex
\begin{minipage}[b]{0.5\linewidth}
\centering
\includegraphics[width=\textwidth]{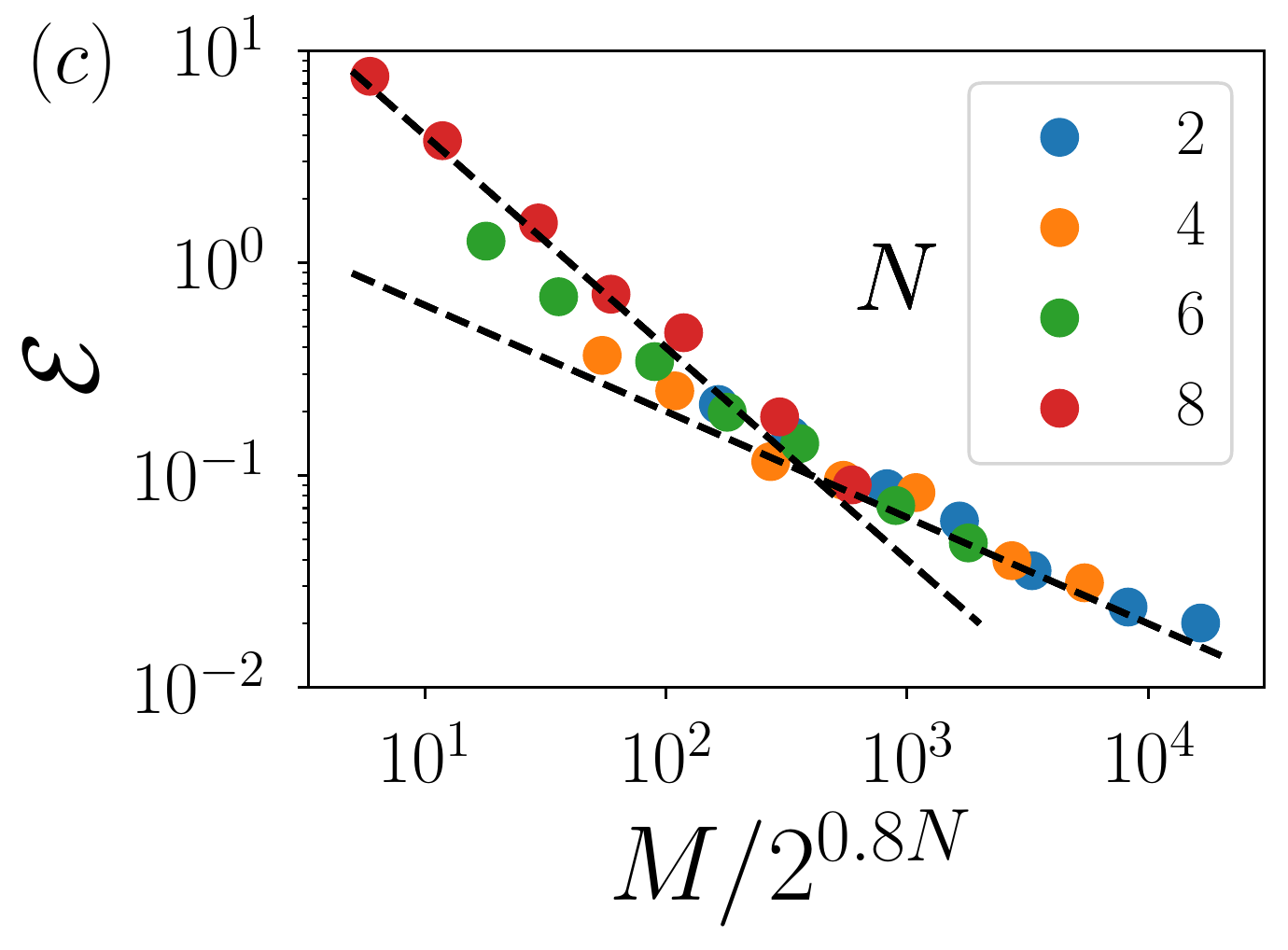}
\end{minipage}
\caption{{\it Protocol and statistical error scaling --} Panel (a) illustrates the randomized measurement protocol needed to estimate the lower bounds of the QFI. Local random unitaries are applied followed by measurements performed in a fixed computational basis. Panels (b) and (c) show the error scaling for $F_0$ and $F_1$ respectively by considering a GHZ state mixed with a depolarisation noise of strength $p = 0.25$, and for various values of $N$ (see legends). The dashed black lines highlight the different error scalings $\propto 1/M$ and $1/\sqrt{M}$.} \label{fig:GHZ}
\end{figure}

A schematic of the protocol is shown in Fig.~\ref{fig:GHZ}(a).
For concreteness, we first consider a system of $N$ qubits, and \bv{discuss the case of collective spin systems further below}. In our protocol, the $N$-qubit quantum state $\rho$ is prepared in the experiment and we apply local random unitaries $u_i$ sampled from the circular unitary ensemble (CUE) (or a unitary 2-design~\cite{Gross2007}). We record as a bitstring $s = s_1, \dots , s_N$ the outcomes of a measurement in a fixed computational basis. This sequence is repeated for $M$ distinct unitaries $u = u_1 \otimes \cdots \otimes u_N$, for which classical bit-strings $s^{(r)}$ with $r = 1, \dots , M$ are stored~\footnote{Note that, to simplify the experimental procedure, it is also possible to write estimators from randomized measurements based on collecting several bit-strings for each random unitary~\cite{Elben2020b}.}.

From this data, we have enough information to reconstruct the density matrix in the limit $M\to \infty$~\cite{Ohliger2013,Elben2018a,Guta2020}. To access directly the function $F_n$, we use the classical shadow formalism~\cite{Huang2020}, and assign to each recorded bit-string $s^{(r)} = s_1^{(r)}, \dots , s_N^{(r)}$ an operator
\begin{equation}
    \hat{\rho}^{(r)} = \bigotimes_{l = 1}^N \Big[ 3\, (u_l^{(r)})^{\dagger} \ket{s_l^{(r)}}\bra{s_l^{(r)}} u_l^{(r)} - \mathbb{1}_2 \Big]. \label{shadow}
\end{equation}
The operator $\hat{\rho}^{(r)}$ is known as a `classical shadow' in the sense that the average over the unitaries and the bit-string measurement results gives $\mathbb{E}[\hat{\rho}^{(r)}]=\rho$.
For different independently sampled shadows labelled $r,r'$, we obtain similarly that $\hat{\rho}^{(r)}\hat{\rho}^{(r')}$ are unbiased estimations of $\rho^2$~\cite{Huang2020}. This approach using $\mathrm{U}$-statistics~\cite{Hoeffding1992} generalizes to estimate any power $\rho^j$, by using $j$ different shadows $\hat{\rho}^{(r_1)},\dots, \hat{\rho}^{(r_j)}$. Then by linearity, we write the unbiased estimator $\hat{F}_n$ of $F_n$ using combinations of $M \geq n+2$ different shadows $\rho^{(r)}$. In particular, for $n=0,1$, from Eqs.~\eqref{F0}--\eqref{F1} we obtain the following unbiased estimators for $F_0$ and $F_1$, respectively:
\bv{
\begin{align}
    \hat{F}_0 & = \frac{4}{2!} \binom{M}{2}^{-1}\sum_{r_1 \ne r_2} \tr \big( \hat{\rho}^{(r_1)}\big[\hat{\rho}^{(r_2)},A\big]A\big) \,, \label{F0-shadow} \\
    \hat{F}_1 & = 2\hat{F}_0 - \frac{4}{3!} \binom{M}{3}^{-1} \!\!\!\!\sum_{r_1 \ne r_2 \ne r_3}\!\!\! \tr \big( \hat{\rho}^{(r_1)}\hat{\rho}^{(r_2)}\big[\hat{\rho}^{(r_3)},A\big]A\big). \label{F1-shadow}
\end{align}}
\ar{Notice that, using independent measurements, the experimental setting of randomized measurements remains the same as in quantum state tomography (QST).}
\ar{To measure a \cb{$N$-qubit} quantum state $\rho$ of rank $\chi$ using QST with $\epsilon-$ accuracy in terms of trace distance, requires $M =O(\chi^2 2^N/\epsilon^2)$ measurements ~\cite{Haah2017}.} 
\ar{Meanwhile for randomized measurements that give access to polynomial functions of $\rho$, the number of measurements to overcome statistical errors} scale as $2^{aN}$ with $a \sim 1$~\cite{Vermersch2019, Elben2018, Elben2020a}. 
\cb{Furthermore,} in tomography the classical post-processing of the measurement data is expensive, as it is based on storing and manipulating exponentially large matrices. In contrast, the use of classical shadows in randomized measurements, which have a tensor product structure, c.f. Eq.~\eqref{shadow}, leads to cheap estimation algorithms in postprocessing runtime and memory usage~\cite{Huang2020,Elben2020b}.

{\it Statistical errors ---} 
Statistical errors associated with the estimation of $F_n$ arise due to the application of a finite number $M$ of random unitary transformations. In particular, as $n$ increases, while the bound $F_n$ becomes tighter the degree of the polynomial in $\rho$ of $\hat F_n$, evaluated with $M$ different shadows, increases.
\ar{In order to provide rigorous performance guarantees for our protocol, we 
analytically study the probabilities $\mathrm{Pr}[|\hat{F}_n - F_n| \geq \epsilon]$ that the statistical errors are larger than a certain accuracy $\epsilon$.}
\ar{Using the Chebyshev's inequality $\mathrm{Pr}[|\hat{F}_n - F_n| \geq \epsilon] \leq \mathrm{Var}[\hat{F}_n] / \epsilon^2$, we relate these to the variance $\var[\hat{F}_n]$ of our estimations. As shown in the SM~\cite{SM}, we provide an upper bound to $\var[\hat{F}_n]$ by generalizing the results of Ref.~\cite{Huang2020,Elben2020b} to arbitrary density matrix polynomials.
This allows us to calculate for any $N-$qubit quantum state $\rho$ the required number of measurements $M$ to estimate $\hat{F}_n$ within a certain confidence interval, so that $\mathrm{Pr}[|\hat{F}_n - F_n| \geq \epsilon] \leq \delta$ for a given $\delta$.
Our results show that the required value of $M$ scales as $\alpha 2^N$ w.r.t $N$, where $\alpha$ can be calculated for any order $n$ based on the knowledge of the state $\rho$ and the operator $A$~\cite{SM}.
As an illustration, we calculate in particular the value of $\alpha$ for pure GHZ states, and in the limit of high accuracy $\epsilon \to 0$. 
We find that, while $F_0$ can be evaluated with fewer measurements compared to $F_1$, there is an overall scaling of $2^N$ for both required values of $M$.
}

To complement our analytical study, we numerically study the error scalings for $F_0$ and $F_1$ by simulating the experimental protocol.
The average statistical error $\mathcal{E}$ is computed by averaging over 50 simulated experimental runs the relative error $\hat{\mathcal{E}} = |\hat{F}_n - F_n|/F_n$ of the estimated bound $\hat{F}_n$. 
We consider the $N$-qubit noisy GHZ state with $A = \frac{1}{2} \sum_{l = 1}^{N} \sigma_z^{(l)}$. 
Fig.~\ref{fig:GHZ}(b,c) show the scaling of the average statistical error in estimating $F_0$ and $F_1$, as a function of rescaled number of measurements $M/2^{aN}$ with $a$ being adjusted by collapsing the data obtained for different $N$ onto a single curve. The figures show that the required number of measurements to obtain an error accuracy of 0.1 scales overall as $\sim 2^{0.7N}$ and $\sim 2^{0.8N}$ for $F_0$ and $F_1$ respectively. 
In particular, for large $M$ we observe a $1/\sqrt{M}$ scaling, which is the standard error decay obtained by performing an empirical Monte-carlo average. In the regime of smaller $M$, the statistical error being high, decays much faster as $1/M$. These two error regimes are also apparent in the expression of the variance (see SM~\cite{SM}). Second, as expected \ar{from the analytical expressions} and seen above from the scaling exponents of $F_0$ and $F_1$, we observe that estimating $F_1$ requires slightly more measurements compared to $F_0$ (see also SM for non-rescaled data). 
Note that similar error scalings, and transition from  $1/\sqrt{M}$ to  $1/M$ behaviors have been observed for other types of cubic order terms related to entanglement negativities~\cite{Zhou2020,Elben2020b}.
\ar{In addition, importance sampling approaches can be incorporated in randomized measurement protocols, leading to 
change of scaling exponents governing statistical errors, and thus drastic reductions of the required number of measurements~\cite{Hadfield2020,Rath2021,Huang2021,Hillmich2021}.}

{\it Systematic errors---}
\ar{Present quantum devices are vulnerable to systematic errors due to noise~\cite{Preskill2018}. 
However, the effect of noise occurring during a measurement can be mitigated. 
First, in the presence of depolarization or qubit readout errors, estimation formulas from randomized measurements can be corrected to provide unbiased estimations ~\cite{Vermersch2018,Elben2020a,Vovrosh2021,Satzinger2021}.}
\bv{Furthermore, in the classical shadows formalism, robust estimations can be performed in the presence of an \emph{unknown} noise channel. This is achieved via a calibration step, which uses a state that can be prepared with high fidelity~\cite{Chen2020,Koh2020,Berg2020}.
Under the assumption of gate-independent, Markovian noise, the data obtained from such calibration provides a model to build robust classical shadows from randomized measurements. 
In the presence of gate-dependent noise, this framework can also be used, showing as well error mitigation~\cite{Chen2020,Koh2020,Berg2020}. These techniques can be applied readily in our protocol.}

{\it Protocol for collective spin systems---}
 We now extend our approach to an ensemble of $N$ particles described by a collective spin \mbox{$S=\frac{N}{2}$}. These systems implemented with ultracold atoms or trapped ions, are relevant to quantum metrology~\cite{ Monz2011,Strobel2014,Barontini2015,Bohnet2016,Garttner2017} as they can feature large-scale multipartite entanglement~\cite{Schmied2016}. Remarkably, our protocol provides access to the series $F_n$ in these systems with relatively low numbers of measurements $M$. 

\begin{figure}[t]
\centering
\begin{minipage}[b]{0.75\linewidth}
\includegraphics[width=\textwidth]{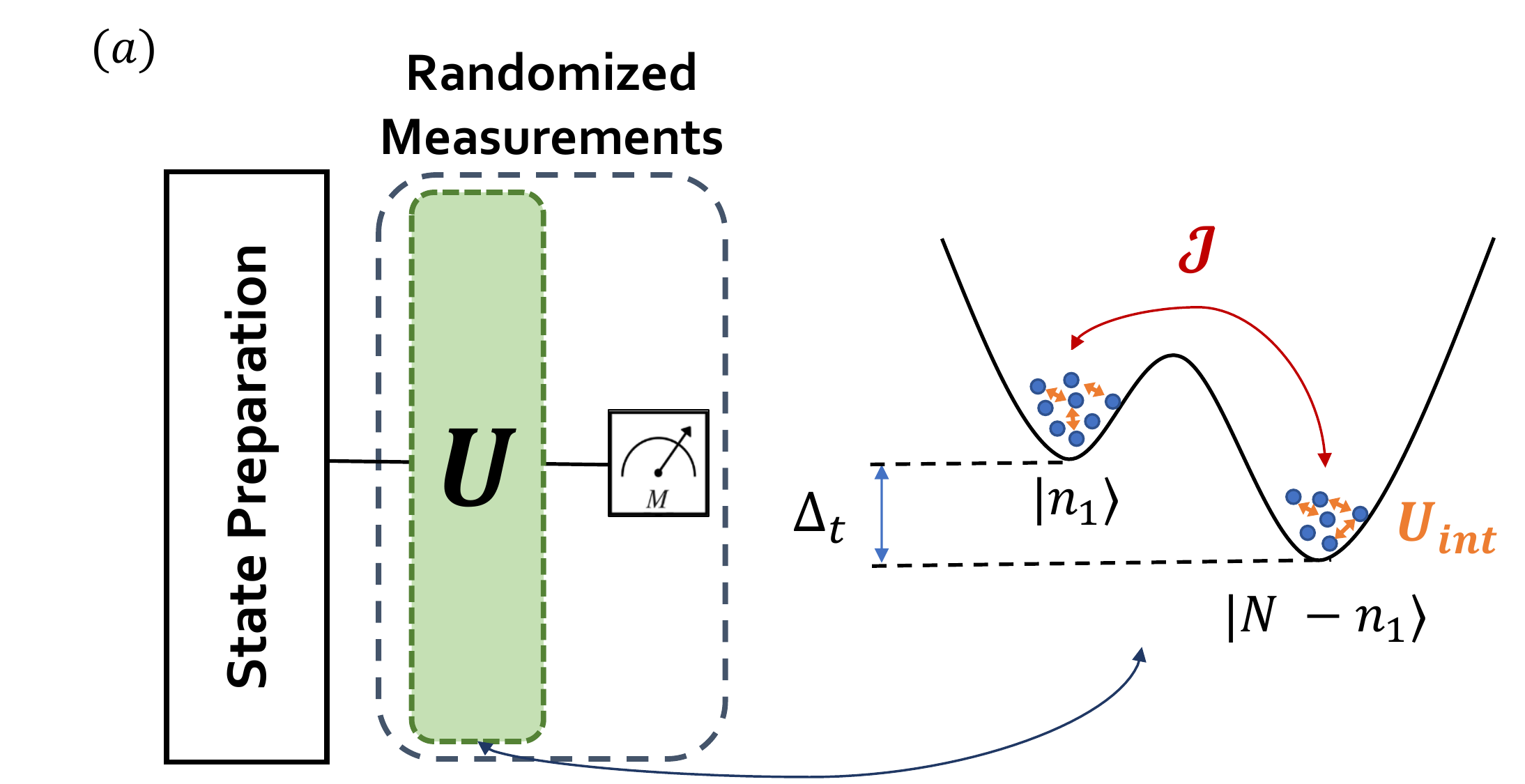}
\end{minipage}
\begin{minipage}[b]{0.5\linewidth}
\centering
\includegraphics[width=\textwidth]{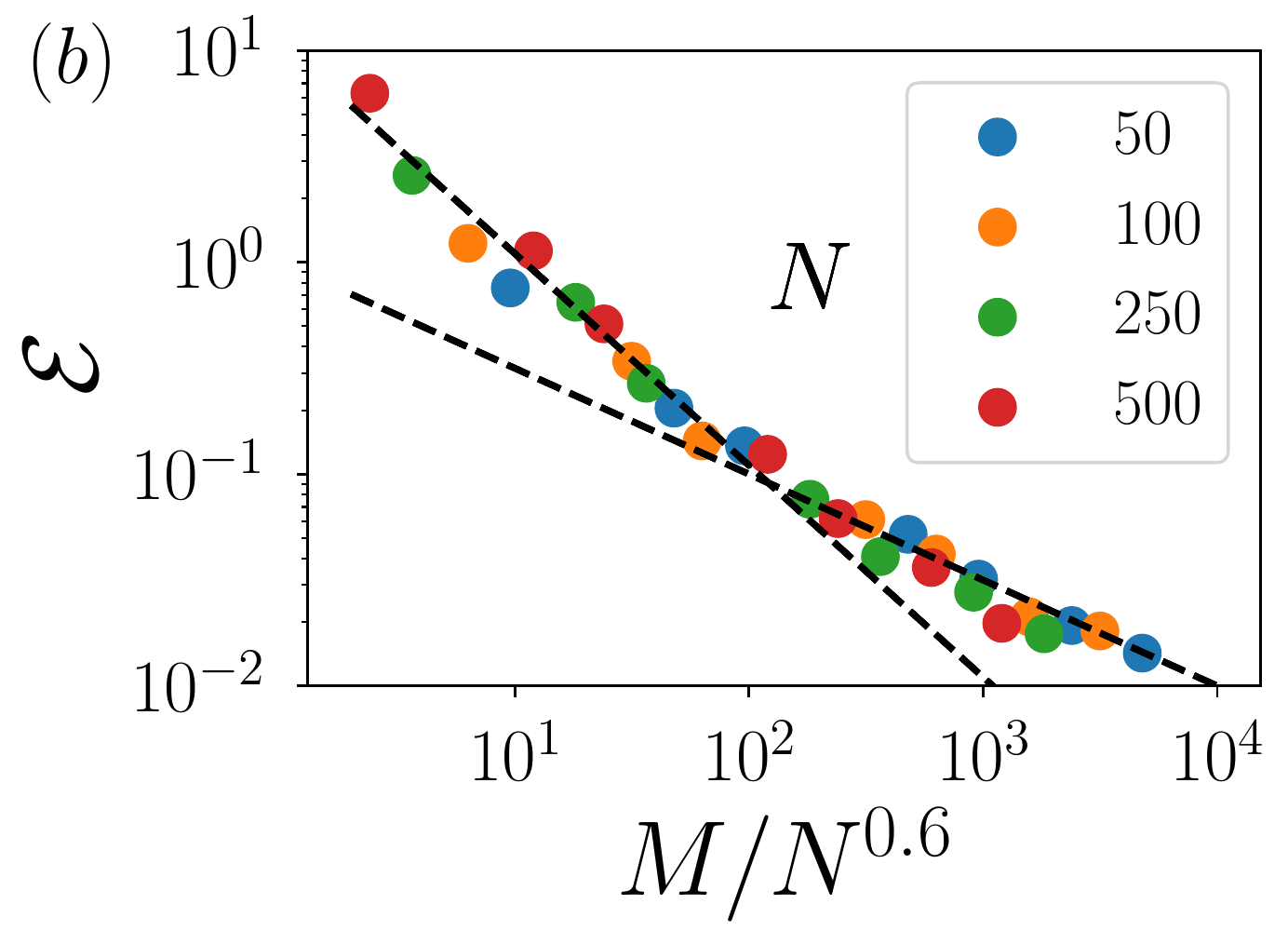}
\end{minipage}
\hskip -1ex
\begin{minipage}[b]{0.5\linewidth}
\centering
\includegraphics[width=\textwidth]{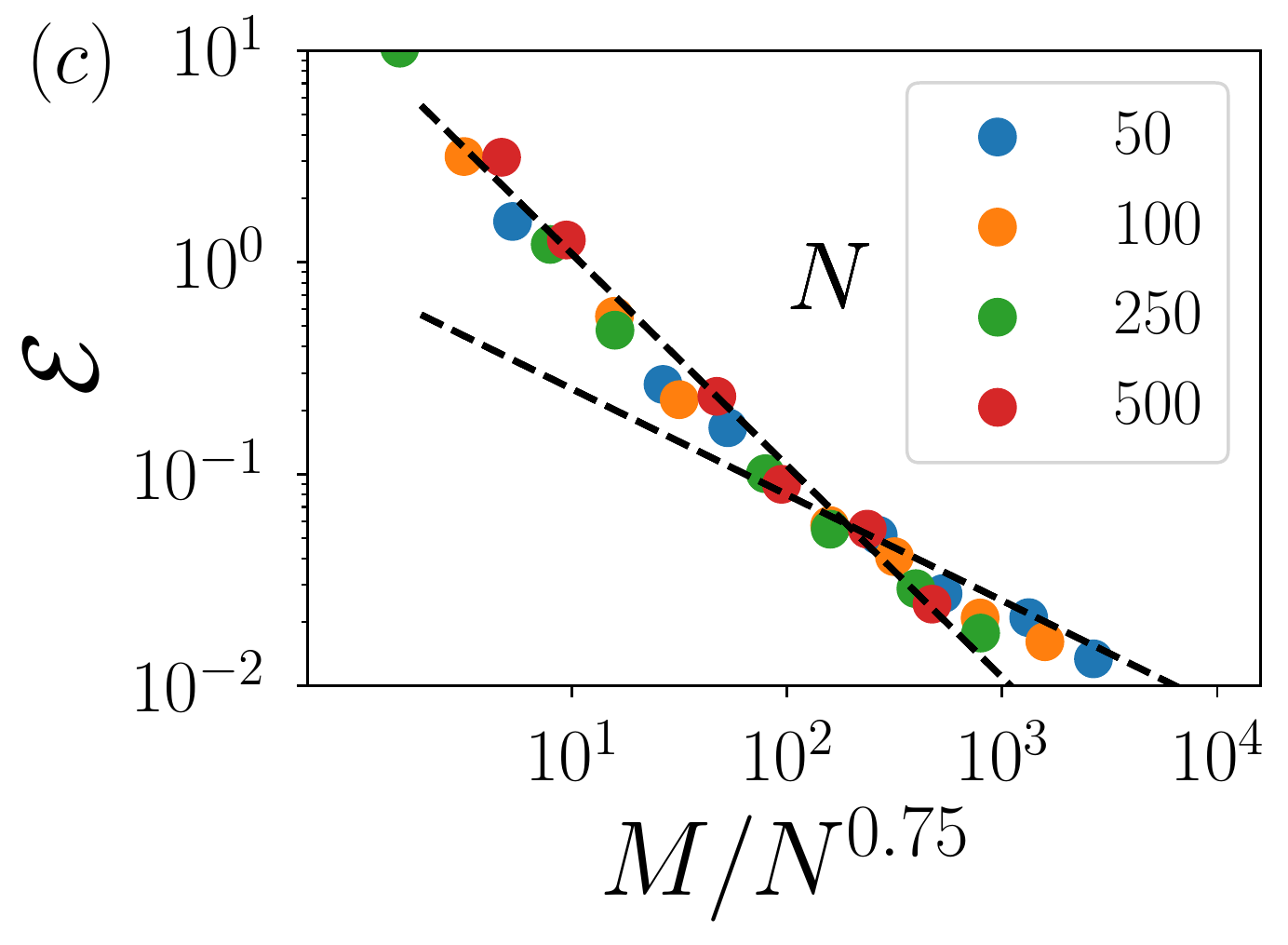}
\end{minipage}
\caption{{\it Protocol and error scalings for collective spin models --} Panel (a) illustrates the two-site collective spin model followed by the randomized measurement protocol implementing global random unitaries. The lower panels show the statistical error scalings of (b) $F_0$ and (c) $F_1$ for a N00N state of bosonic ensembles mixed with depolarisation noise of strength $p = 0.25$ as a function of rescaled axis $M/N^a$, and for various values of $N$ (see legends).\label{fig:NOON}
}
\end{figure}

Consider for concreteness a set of $N$ ultracold bosons in a double-well potential, as illustrated in Fig.~\ref{fig:NOON}(a). It is convenient to write the state of the system in terms of $N+1$ Fock states $\ket{n_1,n_2}$ or $\ket{n_1-n_2}$ with $n_1\in 0,\dots,N$, the number of atoms in the left well and $n_2=N-n_1$ atoms in the right well.  The Bose-Hubbard Hamiltonian $H_t$ describing this system reads
\begin{equation}
    H_t = \frac{\mathcal{J}}{2}(\hat{a}_L^{\dagger}\hat{a}_R + \mathrm{h.c}) + \frac{U_{\mathrm{int}}}{2} \sum_{\ell = L, R} \hat{n}_{\ell}(\hat{n}_{\ell} -1) + \Delta_t(\hat{n}_L - \hat{n}_R), \label{eq:Ht}
\end{equation}
where, $\hat{n}_{L,R} = \hat{a}^{\dagger}_{L,R}\hat{a}_{L,R}$ are the number operators given in terms of creation $\hat{a}^{\dagger}_{L,R}$ and annihilation $\hat{a}_{L,R}$ operators, $\mathcal{J}$ is the tunneling matrix element, $U_\mathrm{int}$ is the on-site interaction energy, and $\Delta_t$ is a random energy offset, c.f Fig.~\ref{fig:NOON}(a). It can be equivalently written using spin $S=\frac{N}{2}$ operators~\cite{arecchi1972atomic,milburn1997quantum,Ferrini2008}. 
The random unitaries $U$ can be experimentally generated as $U=e^{-i H_\eta T} \cdots e^{-iH_1T}$ by choosing different random energy difference $\Delta_\eta$ in $H_\eta$  for some time interval $T$. The convergence of such unitaries $U$ to unitary $2$-designs (which are required in order to build classical shadows) as a function of the depth $\eta$ has been studied in~\cite{Dankert2009,Banchi2017,Elben2018,Vermersch2018,Sieberer2019}. Note that these unitaries $U$, considered here for randomized measurements, can also be used to generate metrologically useful quantum states~\cite{Oszmaniec2016}.

Compared to the situation of $N$ qubits, the protocol to measure the series $F_n$ in this system differs only in applying global random unitaries $U$ instead of local random unitaries $u$. We first prepare a state $\rho$ of interest in this system, then generate and apply random unitaries $U$ followed by measurements of the populations $(n_1,n_2)$ in each well. Repeating this procedure for $M$ unitary matrices $U^{(r)}$, we collect measurement results $s^{(r)}=n_1^{(r)} - n_2^{(r)}$ with $r = 1, \dots , M$. 
A classical shadow  $\hat{\rho}^{(r)}$ of $\rho$~\cite{Ohliger2013,Elben2019,Huang2020} is constructed from each measurement outcome $s^{(r)}$ and the applied unitary $U^{(r)}$ as
\begin{equation}
    \hat{\rho}^{(r)} = \Big[ (N+2)\, (U^{(r)})^{\dagger} \ket{s^{(r)}}\bra{s^{(r)}} U^{(r)} - \mathbb{1}_{N+1} \Big]. 
\end{equation}
Here, the classical shadow is a matrix of dimension \mbox{$(N+1)\times (N+1)$}. From this, we build our series of estimators $\hat{F}_n$ from the measurement data in the same way as shown earlier for the $N$-qubit systems. 

We investigate the statistical error scalings to estimate the lower bounds $F_0$ and $F_1$ by computing the average relative error $\mathcal{E}$. Consider the N00N state given by \mbox{$\ket{\mathrm{N00N}} = \frac{1}{\sqrt{2}}(\ket{N,0} + \ket{0,N})$}~\cite{Pezze2018}. Such states provide optimal metrological sensitivities and are genuinely multipartite entangled, i.e. they have an entanglement depth of $N$~\cite{Pezze2018}. The observable $A$ is taken to be the population difference between the two wells and is defined as $A =  \hat{n}_1 - \hat{n}_2$. We numerically analyze the scaling of statistical errors by directly generating global random unitaries from the CUE.
For simplicity, we study a state \mbox{$\rho(p) = (1-p)\ket{\mathrm{N00N}}\bra{\mathrm{N00N}} + p\mathbb{1}/(N+1)$} subject to depolarisation noise of strength $p = 0.25$. Various other noise models have been studied for macroscopic superposition states in~\cite{huang2006creation,ferrini2010noise,pawlowski2013macroscopic,pawlowski2017mesoscopic}.
\bv{In the case of an \emph{a priori} unknown noise channel, one can use again the formalism of robust shadows to correct our estimates of $F_n$~\cite{Chen2020,Koh2020,Berg2020}.}
Fig.~\ref{fig:NOON}(b,c) show that the required number of measurements here no longer scale exponentially but sub-polynomially in the number of atoms $N$ for $F_0$ and $F_1$. This is attributed to the fact that the Hilbert space dimension scales linearly in $N$. 
The required number of measurements for obtaining an error of 0.1 is found to scale as $N^{a_0}$ and $N^{a_1}$ with $a_0 \sim 0.6$ and $a_1 \sim 0.75$ for $F_0$ and $F_1$, respectively.

{\it Conclusion and outlook ---} 
\bvv{Our method to access the QFI can be used for asserting states capable of providing enhanced metrological sensitivities, or in the context of entanglement detection and quantum simulation. 
Importantly,  we can make predictions on the required number of measurements to detect entanglement with a certain confidence interval.}
As a future direction, it would be interesting to 
compare the entanglement detection `power' of different protocols based on randomized measurements. This includes in particular approaches based on the positive partial transpose condition~\cite{Elben2020b,Yu2021,Neven2021}, which can be applied in the multipartite case~\cite{Jungnitsch2011}, and on statistical correlations of Pauli measurements~\cite{Knips2019,Ketterer2019,Ketterer2020a,Ketterer2020,Imai2021}.

\begin{acknowledgments} 
We thank A. Elben and R. Kueng for fruitful discussions, and comments on the manuscript. BV thanks P. Zoller for inspiring discussions and previous collaborations on randomized measurements.
AR is supported by Laboratoire d'excellence LANEF in Grenoble (ANR-10-LABX-51-01) and from the Grenoble Nanoscience Foundation.
BV acknowledges funding from the Austrian Science Fundation (FWF, P 32597 N). BV and AM acknowledge funding from the French National Research Agency (ANR-20-CE47-0005, JCJC project QRand).
For our numerical simulations we used the quantum toolbox QuTiP~\cite{Johansson2013}.
\end{acknowledgments}

%

\clearpage
\onecolumngrid
\appendix
\begin{center}\LARGE{SUPPLEMENTAL MATERIAL}\end{center}

The Supplemental Material below provides additional elaboration to some of the ideas introduced in the main text. We start with Appendix~\ref{app_proof_fn}, where we detail the proof of our converging polynomial series of lower bounds $F_n$ to the quantum Fisher information (QFI) along with alternate expressions of them; followed by the demonstration in Appendix~\ref{app_convergence} that these bounds converge exponentially to $F_Q$ as $n \to \infty$. Next, we go on to elaborate in Appendix~\ref{app_unbiased_estimates} the construction of unbiased estimators $\hat{F}_n$ of $F_n$ from randomized measurements. In Appendix~\ref{app_general_error_bound_estimation}, we provide the general recipe to evaluate the required number of randomized measurements $M$ to estimate an arbitrary polynomial of the density matrix $\rho$ with a given accuracy $\epsilon$, and a confidence level $\delta$. We follow this by providing an illustration of the same for quadratic and cubic monomials of $\rho$ in Appendix~\ref{app_p2_p3}. Appendix~\ref{app-bound-fn} translates the findings of Appendix~\ref{app_general_error_bound_estimation} to evaluate required number of randomized measurements for any lower bound $F_n$ and we give an illustration of this formalism for $F_0$ and $F_1$ in Appendix~\ref{app_f0_f1}. Finally in Appendix~\ref{app_numerics}, we complement with some additional numerical simulations of our protocol, the analysis of the scalings of the average statistical errors $\cal{E}$ for $F_0$ and $F_1$.

\appendix
\section{Derivation of our lower bounds $F_n$} 
\label{app_proof_fn}
\noindent
In this section we prove that the quantities $F_n$ defined in Eq.~(2) of the main text are indeed lower bounds on the quantum Fisher information.

Let us start by clarifying the form of the QFI in Eq.~(1) of the main text. First note that the denominator in the fraction may not be invertible; in that case the fraction denotes a multiplication by the Moore-Penrose pseudoinverse of the denominator; note also that this (pseudo)inverse commutes with the numerator, so that the fraction notation is indeed nonambiguous.

Consider the spectral decomposition of $\rho$ in the form $\rho = \sum_i \lambda_i \ket{i}\bra{i}$, with $\lambda_i \ge 0$, $\sum_i \lambda_i = 1$ (all the systems and states we consider are finite-dimensional). 
Noting that the numerator in the fraction of Eq.~(1) is $(\rho \otimes \mathbb{1} - \mathbb{1} \otimes \rho)^2 = \big(\sum_{i,j} (\lambda_i \ket{i}\bra{i} \otimes \ket{j}\bra{j} - \lambda_j \ket{i}\bra{i} \otimes \ket{j}\bra{j})\big)^2 = \sum_{i,j} (\lambda_i - \lambda_j)^2 \ket{i,j}\bra{i,j}$ and that the (pseudo)inverse of the denominator is $(\rho \otimes \mathbb{1} + \mathbb{1} \otimes \rho)^{-1} = \big(\sum_{i,j} (\lambda_i \ket{i}\bra{i} \otimes \ket{j}\bra{j} + \lambda_j \ket{i}\bra{i} \otimes \ket{j}\bra{j})\big)^{-1} = \sum_{i,j:\lambda_i + \lambda_j>0} (\lambda_i + \lambda_j)^{-1} \ket{i,j}\bra{i,j}$, we find that Eq.~(1) can be more explicitly written as
\begin{equation}
    F_Q = 2\sum_{\substack{i, j:\\ \lambda_i + \lambda_j>0}} \frac{(\lambda_i - \lambda_j)^2}{\lambda_i + \lambda_j} \tr \big( \ket{i,j}\bra{i,j} \mathbb{S}(A \otimes A) \big) = 2\sum_{\substack{i, j:\\ \lambda_i + \lambda_j>0}} \frac{(\lambda_i - \lambda_j)^2}{\lambda_i + \lambda_j} |\bra{i}A\ket{j}|^2, \label{QFI-SM}
\end{equation}
which is indeed the more common closed form for the QFI that one finds in the literature~\cite{Caves1994,Braunstein1996}.
\medskip

In order to construct lower bounds on the QFI that can suitably be accessed through randomized measurements, the idea is to bound each factor $\frac{(\lambda_i - \lambda_j)^2}{\lambda_i + \lambda_j}$ by a polynomial function of the eigenvalues $\lambda_i$. This can be done by expanding the fraction in a Taylor series (noting that $0 < \lambda_i + \lambda_j \le 1$), and truncating the series as follows:
\begin{equation}
    \frac{1}{\lambda_i + \lambda_j} = \sum_{\ell = 0}^{\infty} (1-\lambda_i - \lambda_j)^\ell \ge \sum_{\ell = 0}^{n} (1-\lambda_i - \lambda_j)^\ell
\end{equation}
for any $n\in\mathbb{N}$.
Thus expressing Eq.~\eqref{QFI-SM} in terms of this expansion gives
\begin{align}
\centering
    &&
    F_Q &= 2\sum_{i, j} \sum_{\ell = 0}^{\infty} (\lambda_i - \lambda_j)^2 (1-\lambda_i-\lambda_j)^\ell |\bra{i}A\ket{j}|^2 \notag\\
    &&
    &= 2 \, \mathrm{Tr} \bigg( \sum_{\ell = 0}^{\infty} (\rho \otimes \mathbb{1} - \mathbb{1} \otimes \rho)^2 (\mathbb{1} \otimes \mathbb{1} -\rho \otimes \mathbb{1} - \mathbb{1} \otimes \rho)^\ell \mathbb{S}(A \otimes A)\bigg) \notag\\ 
    &&
    &\geq 2\sum_{i, j} \sum_{\ell = 0}^{n} (\lambda_i - \lambda_j)^2 (1-\lambda_i-\lambda_j)^\ell |\bra{i}A\ket{j}|^2  \notag\\ 
    &&
    &= 2 \, \mathrm{Tr} \bigg( \sum_{\ell = 0}^{n} (\rho \otimes \mathbb{1} - \mathbb{1} \otimes \rho)^2 (\mathbb{1} \otimes \mathbb{1} -\rho \otimes \mathbb{1} - \mathbb{1} \otimes \rho)^\ell \mathbb{S}(A \otimes A)\bigg) \notag\\
    &&
    &= F_n. \label{qfi-bound-proof}
\end{align}
It is clear from Eq.~\eqref{qfi-bound-proof} that for any observable $A$ and any state $\rho$, $F_n$ monotonically increases ($F_{n+1} \geq F_n$), that $\forall n \in \mathbb{N}$, $F_Q \geq F_n$\, and that $F_n$ converges to $F_Q$ as $n \to \infty$. It is also easy to check that all $F_n$'s are directly equal to $F_Q$ for pure and fully mixed states (in the latter case, $F_n = F_Q = 0$).

\medskip

By further expanding the powers $(\mathbb{1} \otimes \mathbb{1} -\rho \otimes \mathbb{1} - \mathbb{1} \otimes \rho)^\ell = \sum_{q=0}^\ell \binom{\ell}{q} (-1)^q (\rho \otimes \mathbb{1} + \mathbb{1} \otimes \rho)^q$ in Eq.~\eqref{qfi-bound-proof}, swapping the sums and using the {\it hockey-stick identity} $\sum_{\ell = q}^{n} \binom{\ell}{q} = \binom{n+1}{q+1}$, one can express $F_n$ as 
\begin{equation}
    F_n = \sum_{q = 0}^{n} \binom{n+1}{q+1} (-1)^q P_{q+2} \label{QFI-hockey}
\end{equation}
where
\begin{equation}
    P_{q+2} = 2 \, \mathrm{Tr} \Big( (\rho \otimes \mathbb{1} - \mathbb{1} \otimes \rho)^2 ( \rho \otimes \mathbb{1} + \mathbb{1} \otimes \rho)^q \mathbb{S}(A \otimes A)\Big) 
\end{equation}
contains the terms with an order $q+2$ in the density matrix $\rho$.
We can give an alternative expression for $P_{q}$ more directly in terms of powers of $\rho$ and the operator $A$ by expanding the terms inside the trace in the previous expression.
After some manipulation of the sums, and introducing the coefficients $C_m^{(q)} = \binom{q}{m} - 2 \binom{q}{m-1} + \binom{q}{m-2}$, we get%
\footnote{Using the symmetry rule of the binomial coefficients and the cyclic property of the trace, one may note that the summands $C_m^{(q)} \, \tr(\rho^{q+2-m} A\rho^m A)$ are invariant under $m \leftrightarrow (q+2-m)$, so that the sum in Eq.~\eqref{eq_Pq2} actually only contains $\ceil{\frac{q+3}{2}}$ different terms. (Note also that the binomial coefficients appearing in $C_m^{(q)}$ are defined such that $\binom{q}{m'} = 0$ if $m'<0$ or $m'>q$.)}
\begin{equation}
    P_{q+2} = 2 \sum_{m = 0}^{q+2} C_m^{(q)} \, \tr(\rho^{q+2-m} A\rho^m A).
    \label{eq_Pq2}
\end{equation}

\vspace{0.25cm}

Introducing now the cyclic permutation operator \mbox{$\Pi_{(q+2)} = \sum_{i_1,i_2,\ldots,i_{q+2}} \ket{i_{2},\ldots,i_{q+2},i_{1}}\!\bra{i_1,i_2,\ldots,i_{q+2}}$} (where the $\ket{i_m}$'s are basis states for the Hilbert space that contains $\rho$, and with implicit tensor products), which acts on $q+2$ systems and shifts them a single step backwards ($\Pi_{(q+2)}: \ket{\psi_1}\otimes\ket{\psi_2}\otimes\cdots\otimes\ket{\psi_{q+2}} \mapsto \ket{\psi_2}\otimes\cdots\otimes\ket{\psi_{q+2}}\otimes\ket{\psi_1}$), we can write $\tr(\rho^{q+2} A^2) = \tr(A \rho^{q+2} A) = \tr[(\mathbb{1}^{\otimes(q+1)} \otimes A^2) \Pi_{(q+2)} \rho^{\otimes(q+2)}]$ and $\tr(\rho^{q+2-m} A\rho^m A) = \tr[(\mathbb{1}^{\otimes(q+1-m)} \otimes A \otimes \mathbb{1}^{\otimes(m-1)} \otimes A) \Pi_{(q+2)} \rho^{\otimes(q+2)}]$ for $1\le m\le q+1$.%
\footnote{Or for different copies of $\rho$, as we will use in Appendix~\ref{app_general_error_bound_estimation}, $\tr(\rho^{(1)}\rho^{(2)}\cdots\rho^{(q+2)} A^2) = \tr[(\mathbb{1}^{\otimes(q+1)} \otimes A^2) \Pi_{(q+2)} (\rho^{(1)}\otimes\rho^{(2)}\otimes\cdots\otimes\rho^{(q+2)})]$ and $\tr(\rho^{(1)}\cdots\rho^{(q+2-m)} A\rho^{(q+3-m)}\cdots\rho^{(q+2)} A) = \tr[(\mathbb{1}^{\otimes(q+1-m)} \otimes A \otimes \mathbb{1}^{\otimes(m-1)} \otimes A) \Pi_{(q+2)} (\rho^{(1)}\otimes\rho^{(2)}\otimes\cdots\otimes\rho^{(q+2)})]$ for $1\le m\le q+1$.}
Using this and combining Eqs.~\eqref{QFI-hockey} and~\eqref{eq_Pq2}, we get 
\begin{equation}
    F_n = 2 \sum_{q = 0}^{n} \, \binom{n+1}{q+1} (-1)^q \, \sum_{m = 0}^{q+2} C_m^{(q)} \, \tr(\rho^{q+2-m} A\rho^m A) = 2 \sum_{q = 0}^{n} \, \binom{n+1}{q+1} (-1)^q \, \underbrace{\tr[O^{(q+2)} \rho^{\otimes(q+2)}}_{X_{q+2}} ] \label{fn-gen}
\end{equation}
with the $(q+2)$-copy operator
\begin{equation}
    O^{(q+2)} = \Big[ 2 \, (\mathbb{1}^{\otimes(q+1)} \otimes A^2) + \sum_{m = 1}^{q+1} \, C^{(q)}_m \, (\mathbb{1}^{\otimes(q+1-m)} \otimes A \otimes \mathbb{1}^{\otimes(m-1)} \otimes A) \Big] \Pi_{(q+2)}. \label{eq:Oq2}
\end{equation}

\medskip

Let us finally notice, for the sake of completeness, that one can also iteratively calculate each bound $F_n$ from the previous ones and just the highest-order polynomial $P_{n+2}$, by writing
\begin{equation}
    F_n = (-1)^n \Big[ P_{n+2} - \sum_{r = 0}^{n-1} \binom{n+1}{r+1} (-1)^r F_r \Big]
\end{equation}
(which can be proven by using Eq.~\eqref{QFI-hockey} to expand the $F_r$'s above, swapping the sums and using $\sum_{r=q}^{n-1} \binom{n+1}{r+1} \binom{r+1}{q+1} (-1)^r = \binom{n+1}{q+1} (-1)^{n+1}$).
\section{Convergence study of our lower bounds $F_n$}
\label{app_convergence}
\noindent
In this section, we develop more on the convergence of our bounds $F_n$ to $F_Q$ for states that are neither pure, nor fully mixed. We start with
\begin{equation}
    \frac{1}{\lambda_i + \lambda_j} - \sum_{\ell = 0}^{n} (1-\lambda_i - \lambda_j)^\ell = \frac{(1-\lambda_i - \lambda_j)^{n+1}}{\lambda_i + \lambda_j}.
\end{equation}
For any operator $A$, the finite distance $\xi_n$ between $F_Q$ and $F_n$ can then be written as
\begin{equation}
    \xi_n = F_Q - F_n = 2\sum_{i, j: \lambda_i + \lambda_j>0} \frac{(\lambda_i - \lambda_j)^2}{\lambda_i + \lambda_j} (1-\lambda_i - \lambda_j)^{n+1} |\bra{i}A\ket{j}|^2 = O\big(\zeta^n\big) \label{xi},
\end{equation}
where $O$ is the ``big O notation'' and with
\begin{equation}
    \zeta = \max_{i,j:\lambda_i + \lambda_j>0,\lambda_i \neq \lambda_j,\bra{i}A\ket{j}\neq 0} (1 - \lambda_i - \lambda_j).
\end{equation} 
From the expression above, as $n\to\infty$ we thus have an exponential convergence of our bounds $F_n$ to $F_Q$ for any observable $A$ and any state $\rho$.

\begin{figure}[h]
\begin{minipage}[b]{0.37\linewidth}
\centering
\includegraphics[width=\textwidth]{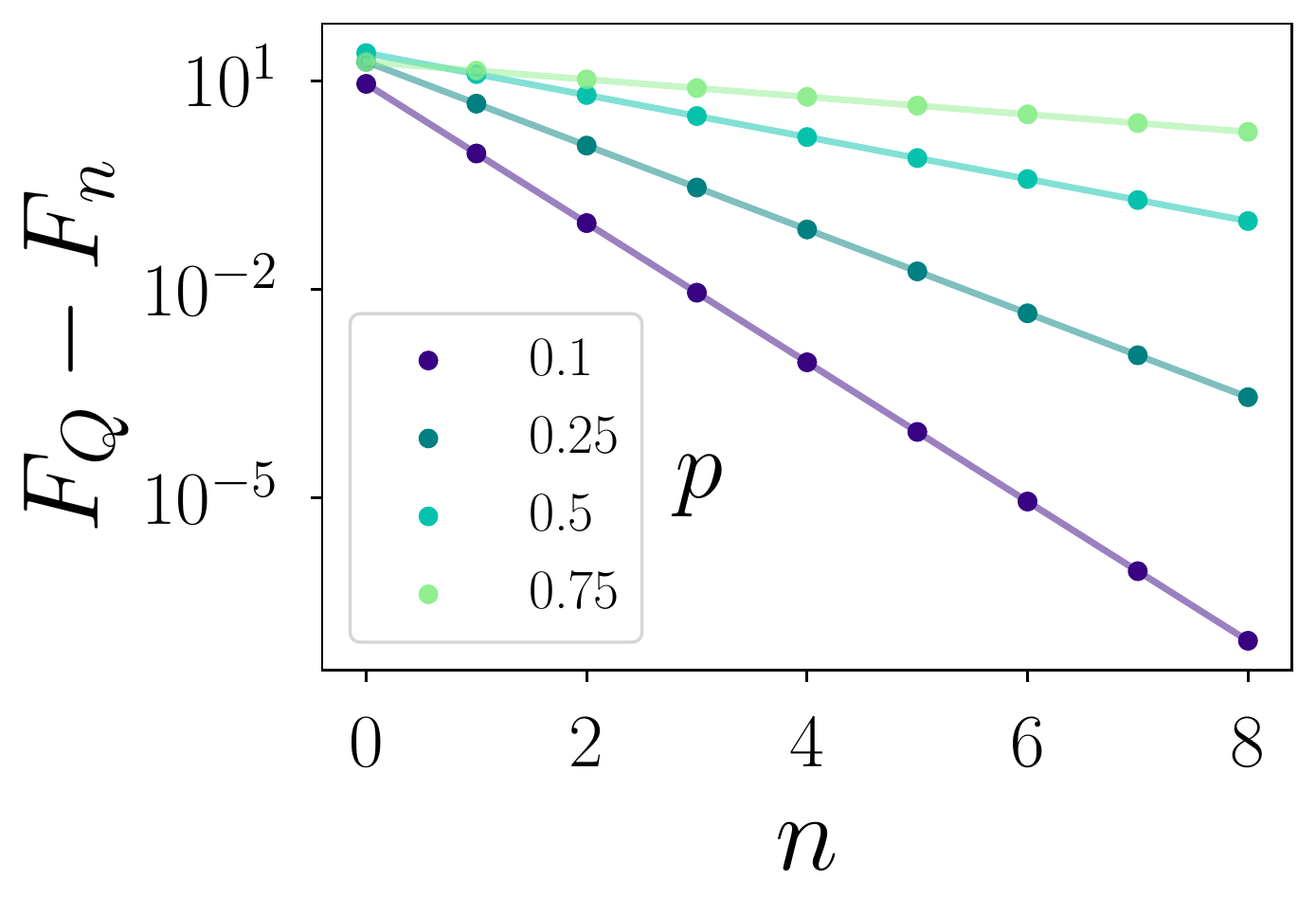}
\end{minipage}
\caption{{\it Exponential convergence of the lower bounds $F_n$ to $F_Q$ ---} The above figure shows, for a noisy 10-qubit GHZ state and $A = \frac{1}{2} \sum_{l = 1}^{N} \sigma_z^{(l)}$, the exponential convergence $F_Q-F_n\propto (1-2^{-9})^n p^n$ for various values of $p$ (see legend). Larger values of $p$ show a slower exponential convergence compared to smaller values of $p$---(as also seen in Fig.~1(a) of the main text).} \label{fig:scaling-convergence}
\end{figure}

To illustrate this point, let us consider a $d$-dimensional pure state $\ket{\psi}$, and define the corresponding state \mbox{$\rho(p) = (1-p)\ket{\psi}\bra{\psi} + p\mathbb{1}/d$}, mixed with the fully mixed state $\mathbb{1}/d$ of noise strength $p$.
Comparing Eqs.~\eqref{QFI-SM} and~\eqref{xi}, and noting that all terms $(1-\lambda_i - \lambda_j)$ appearing in the sum of Eq.~\eqref{xi} are equal to $p + 2p/d$ for such a state, we can express the finite distance $\xi_n$ for $\rho(p)$ as
\begin{equation}
    \xi_n = F_Q(1-2/d)^{n+1}p^{n+1} = (F_Q-F_0) (1-2/d)^n p^n \label{convergence}
\end{equation}
(while the explicit calculation of $F_Q$ gives $F_Q = 4(\bra{\psi}A^2\ket{\psi}-\bra{\psi}A\ket{\psi}^2)\frac{(1-p)^2}{1-p+2p/d}$).
The above expression enables us to see the exponential convergence ($F_Q-F_n \propto (1-2/d)^n p^n$) of the lower bound series $F_n$ to $F_Q$ for any fixed $d-$dimensional state $\rho(p)$ of noise strength of $p$. In particular, we highlight this convergence feature of our bounds in Fig.~\ref{fig:scaling-convergence} by considering noisy GHZ states ($\ket{\psi} = \ket{\mathrm{GHZ}_N}$) with different values of the noise strength $p$, and with respect to the collective spin observable $A = \frac{1}{2} \sum_{l = 1}^{N} \sigma_z^{(l)}$.
\section{Constructing unbiased estimators of our lower bounds $F_n$}
\label{app_unbiased_estimates}
\noindent
In this section, we elaborate on the construction of unbiased estimators of our lower bounds (such that $\mathbb{E}[\hat{F_n}] = F_n$) from the properties of U-statistics~\cite{Hoeffding1992}. To review the protocol introduced in the main text, we perform randomized measurements on a prepared $N-$qubit state by applying local random unitaries $u = u_1 \otimes \dots \otimes u_N$ sampled from the CUE or a unitary 2-design followed by fixed computational basis measurements. Each distinct applied unitary \mbox{$u^{(r)} = u_1^{(r)} \otimes \dots \otimes u_N^{(r)}$} is followed by a computational basis measurement giving the bit-string outcome $s^{(r)} = s_1^{(r)},\ldots,s_N^{(r)}$ with $r = 1, \dots , M$. We define a  single snapshot of the state {\it classical shadow}~\cite{Huang2020} as given previously in Eq.~(5) as
\begin{equation}
    \hat{\rho}^{(r)} = \bigotimes_{l=1}^N \Big[ 3\, (u_l^{(r)})^{\dagger} \ket{s_l^{(r)}}\bra{s_l^{(r)}} u_l^{(r)} - \mathbb{1}_2 \Big], \label{shadow-SM}
\end{equation}
with the classical shadow giving the state in expectation, i.e $\mathbb{E}[\hat{\rho}^{(r)}] = \rho$.

To obtain estimators of $F_n$, we require unbiased estimators of all $X_{q+2} = \tr[O^{(q+2)} \rho^{\otimes(q+2)} ]$ terms in Eq.~\eqref{fn-gen} with  $q = 0, \dots, n$. We note that each term $X_{q+2}$ contains $q+2$ copies of $\rho$ or equivalently a $(q+2)$-order functional in $\rho$. Using U-statistics we get estimators $\hat{X}_{q+2}$ of $X_{q+2}$ by replacing $\rho$, in each of its $q+2$ copies, by $\hat{\rho}^{(r_i)}$ with $i = 1, \dots , q+2$ and $\forall i\neq j, r_i \ne r_j$, corresponding to independent realisations of the single random unitaries $u^{(r_1)}, \dots , u^{(r_{q+2})}$. We can then write the estimator as
\begin{equation}
    \hat{X}_{q+2} = \frac{1}{(q+2)!} \binom{M}{q+2}^{-1} \sum_{r_1 \ne \dots \ne r_{q+2}} \tr \big( O^{(q+2)} \bigotimes_{i = 1}^{q+2} \hat{\rho}^{(r_i)} \big).  \label{bounds-estimators}
\end{equation}
The properties of U-statistics guarantee that $\hat{X}_{q+2}$ is an unbiased estimator of $X_{q+2}$ i.e $\mathbb{E}[\hat{X}_{q+2}] = X_{q+2}$. From this we obtain the final expression of the estimator, using Eq.~\eqref{fn-gen}, as
\begin{equation}
    \hat{F}_n = 2 \sum_{q = 0}^{n} \, \binom{n+1}{q+1} (-1)^q \, \hat{X}_{q+2}. \label{fn-xq}
\end{equation}

\section{Recipe to estimate error bounds and the required number of randomized measurements \\ for an arbitrary order polynomial function of $\rho$}
\label{app_general_error_bound_estimation}
In this section we introduce a method to analytically calculate the error bounds for any order polynomial function of the density matrix $\rho$ describing the state of a $N-$qubit system. This will enable us to estimate the required number of randomized measurements $M$ to estimate a function to be $\epsilon$ close with a confidence level of $\delta$, consequently giving us the measurement cost to estimate any arbitrary order lower bound $F_n$ introduced earlier.
Without loss of generality, we consider a polynomial function $X_q$ generically defined as
\begin{equation}
    X_q = \tr(O^{(q)} \rho^{\otimes q}),
\end{equation}
where the operator $O^{(q)}$ acts on $q$ copies of the density matrix $\rho$ and its U-statistic estimator $\hat{X}_q$ is given by:
\begin{equation}
    \hat{X}_{q} = \frac{1}{q!} \binom{M}{q}^{-1} \sum_{r_1 \ne \dots \ne r_{q}} \tr \big( O^{(q)} \bigotimes_{i = 1}^{q} \hat{\rho}^{(r_i)} \big).
\end{equation}
We intend to calculate the required number of measurements $M$ to estimate $X_q$ with an error $|\hat{X}_q - X_q| \leq \epsilon$ and with a certain confidence level $\delta$. We start by recalling the Chebyshev's inequality that bounds the probability our estimations $\hat{X}_q$ being in a confidence interval:
\begin{equation}
    \mathrm{Pr}[|\hat{X}_q - X_q| \geq \epsilon] \leq \frac{\mathrm{Var}[\hat{X}_q]}{\epsilon^2}. \label{Cby}
\end{equation}
In the rest of this section, we develop the method to bound $\var[\hat{X}_q]$ from above as the convergence of the estimator $\hat{X}_q$ to its true value i.e $X_q = \expp[\hat{X}_q]$ is governed by its variance. Some special cases of variance involving quadratic and cubic single polynomial functions of $\rho$ have been bounded in some previous works~\cite{Huang2020,Elben2020b}. Here, we extend these calculations by providing tighter bounds that could help in the future to estimate the randomized measurement budget of other polynomial functions of $\rho$ of arbitrary order.

\medskip

The variance of $\hat{X}_q$ writes 
\begin{align}
    &&
    \var[\hat{X}_q] &= \expp[\hat{X}_q^2] - \expp[\hat{X}_q]^2 =  \frac{1}{q!^2} \binom{M}{q}^{-2} \sum_{r_1 \ne \dots \ne r_{q}}
     \sum_{r_1'\ne \dots \ne r_{q}'}
     \mathbb{E}\Big[\tr \big( O^{(q)} \bigotimes_{i = 1}^{q} \hat{\rho}^{(r_i)} \big)\tr \big( O^{(q)} \bigotimes_{i = 1}^{q} \hat{\rho}^{(r'_i)} \big)\Big]- X_q^2 \label{var-xq-compact} 
    \\
    && 
    &= \frac{1}{q!^2} \binom{M}{q}^{-2} \sum_{r_1 \ne \dots \ne r_{q}}
     \sum_{r_1'\ne \dots \ne r_{q}'}
     \cov \Bigg[\tr \big( O^{(q)} \bigotimes_{i = 1}^q \hat{\rho}^{(r_i)}\big), \tr \big(O^{(q)} \bigotimes_{i = 1}^{q} \hat{\rho}^{(r'_i)} \big)\Bigg].
     \label{var-xq-compact2}
\end{align}
To compute the above sums over two independent $q$-uplets of indices $\{r_1, r_2, \dots , r_{q}\}$ and $\{r'_1, r'_2, \dots , r'_{q}\}$, one can first sum over the number $k$ of indices that are in common in the two $q$-uplets, then sum over these indices (renaming these joint indices $s_i$) and those that are not in common (renaming these $t_j, t_j'$); in addition, one can take these indices to be ordered, and sum over their permutations. Hence, one can write the previous sums as
\begin{align}
    \var[\hat{X}_q] 
     &=\frac{\binom{M}{q}^{-2}}{q!^2}
     \sum_{k=0}^q
     \sum_{\substack{s_1< \dots < s_k\\
     \neq t_{k+1}< \dots < t_{q}\\
    \neq t'_{k+1}< \dots < t'_{q}}}
     \!\!\sum_{\pi,\tau}
     \cov \Big[
     \tr \big( O^{(q)} \pi[\otimes_{i=1}^k \hat{\rho}^{(s_i)} \otimes_{j=k+1}^q \hat{\rho}^{(t_j)}]\pi^\dag \big),
      \tr \big( O^{(q)} \tau[\otimes_{i=1}^k \hat{\rho}^{(s_i)} \otimes_{j=k+1}^q \hat{\rho}^{(t'_j)}] \tau^\dag \big)
     \Big]
\end{align}
where the $s$ and $t^{(\prime)}$ indices satisfy $s_i \neq t_j, s_i \neq t_j', t_i \neq t_j'$ for all $i,j$, and where $\pi,\tau$ are permutation operators between $q$ copies (generalized swap operators) defined as%
\footnote{With some slight abuse of notations, here we use the same symbol $\pi$ to denote the permutation itself ($j \mapsto \pi(j)$) and the corresponding operator.}
\mbox{$\pi \equiv \sum_{i_1,\ldots, i_q}\ket{i_{\pi(1)},\ldots, i_{\pi(q)}}\bra{i_1,\ldots,i_q}$} and equivalently for $\tau$.
Since each $t_j$ or $t_j'$ index appears only once in the expectation values that compose the covariances, the corresponding $\hat{\rho}^{(t_j^{(\prime)})}$ all simply average to $\rho$; furthermore, all different combinations of the $s$ indices give the same value, equal e.g. to that obtained for $s_i = i \ \forall\, i$. Counting the number of different tuples $s_1< \dots < s_k$, $t_{k+1}< \dots < t_{q}$ and $t'_{k+1}< \dots < t'_{q}$ (which thus all give the same contribution), we get
\begin{align}
    \var[\hat{X}_q] 
     &= \frac{\binom{M}{q}^{\!\!-2}}{q!^2}
     \!\!
     \sum_{k=0}^q
     \!
    \binom{M}{k} \binom{M-k}{q-k} \binom{M-q}{q-k}
    \!\sum_{\pi,\tau}
     \cov \Big[
     \tr \big( \pi^\dag O^{(q)} \pi[\otimes_{i=1}^k \hat{\rho}^{(i)} \rho^{\otimes(q-k)}] \big),
      \tr \big(\tau^\dag O^{(q)} \tau[\otimes_{i=1}^k \hat{\rho}^{(i)}\rho^{\otimes(q-k)}] \big)
     \Big] \nonumber \\[1mm]
     &= \binom{M}{q}^{\!-2}
     \sum_{k=0}^q
    \binom{M}{k} \binom{M-k}{q-k} \binom{M-q}{q-k}
    \var \Big[
     \tr \big( O_k^{(q)} \otimes_{i=1}^k \hat{\rho}^{(i)} \big)
     \Big] \label{eq:VarXq}
\end{align}
with%
\footnote{One may note that the calculation of the operator $O_k^{(q)}$ can be simplified to a calculation of $\binom{q}{k}$ distinct rearrangements with each of the terms occurring $\frac{q!}{\binom{q}{k}}$ times, and the partial trace being taken over the corresponding positions of $\rho$.}
\begin{equation}
    O_k^{(q)}=\frac{1}{q!}\sum_{\pi}\mathrm{Tr}_{\{k+1\dots q\}}(\pi^\dag O^{(q)} \pi[\mathbb{1}^{\otimes k} \otimes \rho^{\otimes (q-k)}]) = \mathrm{Tr}_{\{k+1\dots q\}}(\bar{O}^{(q)}[\mathbb{1}^{\otimes k} \otimes \rho^{\otimes (q-k)}]),\label{oqk} 
\end{equation}
where we defined the ``fully-scrambled'' operator $\bar{O}^{(q)}= \frac{1}{q!}\sum_{\pi} \pi^\dag O^{(q)} \pi$ for later convenience.
Note that for $k =0$, the above variance term is zero (so that the sum can be taken to start at $k=1$). For $k\ge 1$, we notice that the operator $O^{(q)}_k$ acts on $k$ copies of the shadows $\otimes_{i=1}^k \hat{\rho}^{(i)}$. Let us recall some results already proved in previous works~\cite{Huang2020, Elben2020b}, that bound the variance of the single snapshot for a linear function $\tr(\widetilde{O}_1\hat{\rho}^{(1)})$ (D7 in~\cite{Elben2020b}) and quadratic function $\tr(\widetilde{O}_2 \hat{\rho}^{(1)} \otimes \hat{\rho}^{(2)})$ (D12 in~\cite{Elben2020b}) with $\widetilde{O}_1$ and $\widetilde{O}_2$ being operators acting on the single and two copies of the distinct shadows $\hat{\rho}^{(1)} \ne \hat{\rho}^{(2)}$. These bounds can be written as: 
\begin{align}
    \mathrm{Var}\big[\tr\big(\widetilde{O}_1\hat{\rho}^{(1)}\big)\big] & \leq 2^N \tr\big(\widetilde{O}_1^2\big), \label{1-copy-bound} \\
    \mathrm{Var}\Big[\tr\big(\widetilde{O}_2(\hat{\rho}^{(1)} \otimes \hat{\rho}^{(2)})\big)\Big] & \leq 2^{2N} \tr\big(\widetilde{O}_2^{2}\big). \label{2-copy-bound-1}
\end{align}
A generalisation of the above conditions for the variance of a $k-$order function $\tr(\widetilde{O}_k \hat{\rho}^{(1)} \otimes \dots \otimes \hat{\rho}^{(k)})$ where $\widetilde{O}_k$ acts on $k$ copies of shadows with $\hat{\rho}^{(1)} \ne \dots \ne \hat{\rho}^{(k)}$, can be formulated by viewing the $k-$tensor product operator $\bigotimes_{i = 1}^{k} \hat{\rho}^{(i)}$ as a single shadow in an augmented Hilbert space dimension of $2^{kN}$. This allows us to formulate the general bound:
\begin{equation}
    \var\Big[\tr\big(\widetilde{O}_k(\hat{\rho}^{(1)} \otimes \dots \otimes \hat{\rho}^{(k)})\big)\Big] \leq 2^{kN} \tr\big(\widetilde{O}_k^{2}\big). \label{k-copy-bound}
\end{equation}
This general condition enables us to bound $\var[\hat{X}_q]$ from Eq.~\eqref{eq:VarXq} as
\begin{align}
    &&
    \var[\hat{X}_q]   \le
 \sum_{k=1}^q
    \frac{q!^2(M-q)!^2 \,2^{kN}}{M!k!(q-k)!^2(M-2q+k)!}
     \tr \big( [O_k^{(q)}]^2\big)
     \leq 
    \sum_{k=1}^q
    \frac{q!^2 2^{kN}}{k!(q-k)!^2(M-k+1)^k}
     \tr \big( [O_k^{(q)}]^2\big) \label{var-xq}
\end{align}
where, in order to isolate the dependency on $M$ (as ultimately we want to obtain a bound on $M$), we used the fact that $\frac{(M-q)!^2}{M!(M-2q+k)!} = \frac{(M-q)(M-q-1)\ldots(M-2q+k+1)}{M(M-1)\ldots(M-q+1)} = \frac{1}{M(M-1)\ldots(M-k+1)} \frac{M-q}{M-k}\frac{M-q-1}{M-k-1}\cdots\frac{M-2q+k+1}{M-q+1} \le \frac{1}{(M-k+1)^k}$.
The bound on $\var[\hat{X}_q]$ can be viewed as a sum of $q$ different orders of contributions $k$ to the variance. Recalling the Chebyshev’s inequality mentioned in Eq.~\eqref{Cby}, we write 
\begin{equation}
    \mathrm{Pr}[|\hat{X}_q - X_q| \geq \epsilon] \leq \frac{\mathrm{Var}[\hat{X}_q]}{\epsilon^2} \leq \frac{1}{\epsilon^2}\sum_{k = 1}^{q} \frac{q!^2 2^{kN}\tr([O_k^{(q)}]^2)}{k!(q-k)!^2(M-k+1)^k}. \label{var-xq_bnd_M}
\end{equation}
One can ensure that $\mathrm{Pr}[|\hat{X}_q - X_q| \geq \epsilon] \leq \delta$, for a given $\delta \in [0, \,1]$, if for instance each term of the sum above is less than $\delta/q$. Then it follows that in order to assert $X_q$ with a confidence level of at least $1-\delta$ and $|\hat{X}_q - X_q| \leq \epsilon$ for any $\delta,\,\epsilon > 0$, it suffices to take a number of measurements
\begin{equation} 
    M \ge \max_{1 \leq k\leq q }\Bigg\{\Bigg(\frac{q \, q!^2 }{k!(q-k)!^2} \frac{\tr([O_k^{(q)}]^2)}{\epsilon^2 \delta}  \Bigg)^{\frac{1}{k}} 2^{N}+k-1\Bigg \}. \label{conf-xq}
\end{equation}
The above expression provides the required measurement budget $M$ to evaluate the estimator of an arbitrary order polynomial function with any defined accuracy $\epsilon$ and confidence level of $1-\delta$.
In particular, for each of the $q$ possible values of $k$ the number of measurements is proportional to $ \epsilon^{-2/k}$.
In the limit of $\epsilon\to 0$ (and for a fixed $\delta$), the $\max$ in the above equation corresponds to $k=1$ (linear contribution to the variance), with a  required number of measurements $M\propto \epsilon^{-2}$.

\section{Illustration of error bounds and the required number of measurements \\ for quadratic and cubic monomials of $\rho$}
\label{app_p2_p3}
This section is intended for interested readers and serves as an additional illustration. Based on our general expressions of Eq.~\eqref{var-xq} or~\eqref{conf-xq}, we are able to correct and provide improvements on the error bounds previously given in~\cite{Elben2020b} for quadratic and cubic monomials of the state $\rho$, i.e. $p_2 = \tr(\rho^2) = \tr(\Pi_{(2)}\rho^{\otimes 2})$ and $p_3 = \tr(\rho^3) =\tr(\Pi_{(3)}\rho^{\otimes 3})$.
The goal is to calculate the operator $O^{(q)}_k$ defined in Eq.~\eqref{oqk} and the rest of the steps are quite straightforward calculation as shown in the previous section.

For $p_2$, we have $q = 2$ and need to compute the operators $O^{(2)}_k$ with $k = 1,2$ with $O^{(2)} = \Pi_{(2)}$, the two copy swap operator.
The two permutations $\pi$ to sum over in Eq.~\eqref{oqk} are ``the identity" and ``the swap", and it is easily checked that one gets $\bar{O}^{(2)} = \frac12\sum_\pi \pi^\dagger \Pi_{(2)} \pi = \Pi_{(2)}$, so that we get
\begin{eqnarray}
    O^{(2)}_1 = \tr_{\{2\}}(\Pi_{(2)}[\mathbb{1} \otimes \rho]) = \rho\,, \qquad O^{(2)}_2 = \Pi_{(2)}\,,
\end{eqnarray}
so that
\begin{eqnarray}
    \tr([O^{(2)}_1]^2) = \tr( \rho^2)=p_2\,, \qquad \tr([O^{(2)}_2]^2) = \tr(\Pi_{(2)}^2) = 2^{2N}\,.
\end{eqnarray}

Now we can bound the variance of $\hat{p}_2$ by using Eq.~\eqref{var-xq} as
\begin{equation}
    \var[\hat{p}_2] \leq 4\frac{2^N}{M}p_2 + 2\frac{2^{4N}}{(M-1)^2}
\end{equation}
Using Eq.~\eqref{conf-xq} and the above findings, we can ensure that the estimation $\hat{p}_2$ of the purity satisfies \mbox{$\mathrm{Pr}[|\hat{p}_2 - p_2| \geq \epsilon] \leq \delta$} for a given $\delta$ and $\epsilon$ when%
\footnote{In comparison to the expressions of Eqs.~(D16) and~(D17) in Ref.~\cite{Elben2020b}, the above bound provides an improvement along with a correction. Accounting the correction (noting that $\tr(\Pi_{AB}^2)$ in Eq.~(D16) should have been evaluated as $2^{2|AB|}$ instead of $2\times2^{|AB|}$), the expression of (D17) in Ref.~\cite{Elben2020b} should read $M \ge \max\Big\{8 \frac{2^{|AB|}p_2 }{\epsilon^2 \delta}, 2 \frac{2^{2|AB|} }{\epsilon \sqrt{\delta}}\Big\}$ (ignoring here, as in Ref.~\cite{Elben2020b}, the negligible (for large $M$) `+1' term from Eq.~\eqref{eq:M_p2}).} 
\begin{equation}
    M \ge \max\Bigg\{8\,\frac{p_2 2^{N} }{\epsilon^2 \delta}, 2\,\frac{2^{2N} }{\epsilon \sqrt{\delta}}+1\Bigg \}. \label{eq:M_p2}
\end{equation}

Similarly for $p_3$, we start by defining the operator $O^{(3)} = \Pi_{(3)}$. Averaging over the 6 three-copy permutations $\pi$, it is easily found that $\bar{O}^{(3)} = \frac{1}{3!}\sum_\pi \pi^\dagger \Pi_{(3)} \pi = \frac12(\Pi_{(3)} + \Pi_{(3)}^2)$, which leads to
\begin{eqnarray}
    & O^{(3)}_1 = \tr_{\{2,3\}}\big(\frac12(\Pi_{(3)} + \Pi_{(3)}^2)[\mathbb{1} \otimes \rho^{\otimes 2}]\big) = \rho^2\,, \notag \\[1mm]
    & O^{(3)}_2 = \tr_{\{3\}}\big(\frac12(\Pi_{(3)} + \Pi_{(3)}^2)[\mathbb{1}^{\otimes 2} \otimes \rho]\big) = \frac12\big(\rho \otimes \mathbb{1} + \mathbb{1} \otimes \rho\big)\Pi_{(2)}\,, \qquad O^{(3)}_3 = \frac12(\Pi_{(3)} + \Pi_{(3)}^2)\,,
\end{eqnarray}
so that
\begin{eqnarray}
    \tr([O^{(3)}_1]^2) = \tr( \rho^4)\,, \quad \tr([O^{(3)}_2]^2) = \frac12\big(\tr(\rho)^2 + p_2\, 2^N\big) \le p_2\, 2^N\,, \quad \tr([O^{(3)}_3]^2) = \frac12\big(2^{3N}+2^N\big) \le 2^{3N}
\end{eqnarray}
(where in the second expression we used the fact that the purity of a $2^N$-dimensional state satisfies $p_2\, \ge \tr(\rho)^2 / 2^N$).
This enables us to write down the variance bound on $\hat{p}_3$ following Eq.~\eqref{var-xq} as:
\begin{equation}
    \var[\hat{p}_3] \leq  9\frac{2^N}{M}\tr(\rho^4) + 18\frac{2^{3N}}{(M-1)^2}p_2 + 6\frac{2^{6N}}{(M-2)^3}.
\end{equation}
Now, again using Eq.~\eqref{conf-xq}, we ensure that the estimation $\hat{p}_3$  of $p_3$ satisfies $\mathrm{Pr}[|\hat{p}_3 - p_3| \geq \epsilon] \leq \delta$  for a given $\delta$ and $\epsilon$ when%
\footnote{In comparison with expressions of Eqs.~(D28) and~(D29) of Ref.~\cite{Elben2020b}, we improve and provide the corrected bounds. Taking into account the corrections: the expression of (D28) reads $\var[\hat{p}_3] \leq 9\frac{2^{|AB|}}{M}\tr(\rho^4) + 18\frac{2^{3|AB|}}{(M-1)^2}p_2 + 6\frac{2^{6|AB|}}{(M-2)^3}$ and consequently (D29) should read $M\ge \,\max \Big\{27\frac{2^{|AB|}\tr(\rho^4)}{\epsilon^2\delta}, \sqrt{54}\frac{2^{1.5|AB|}\sqrt{p_2}}{\epsilon\sqrt{\delta}}, \sqrt[3]{18}\frac{2^{2|AB|}}{\epsilon^{\frac{2}{3}} \delta^{\frac{1}{3}}}\Big\}$.} 

\begin{equation}
    M \ge \max \Bigg\{27\,\frac{\tr(\rho^4) \,2^{N} }{\epsilon^2 \delta} , \sqrt{54}\,\frac{\sqrt{p_2}\, 2^{3N/2} }{\epsilon \sqrt{\delta}} +1, \sqrt[3]{18}\,\frac{ 2^{2N} }{\epsilon^{\frac{2}{3}} \delta^{\frac{1}{3}}} +2\Bigg \}.
\end{equation}


\section{Error bounds and the required number of measurements \\for arbitrary order lower bounds $F_n$}
\label{app-bound-fn}
Now based on the previous results of Appendix~\ref{app_general_error_bound_estimation}, we can easily translate them to estimate the error bound for an arbitrary lower bound $F_n$ on the quantum Fisher information.
Recalling Eq.~\eqref{fn-xq} and using the Cauchy–Schwarz inequality, we can write
\begin{align}
\var[\hat{F}_n] &= \var\Bigg[2 \sum_{q = 0}^{n} \, \binom{n+1}{q+1} (-1)^q \, \hat{X}_{q+2}\Bigg] \le 4 \Bigg[\sum_{q = 0}^{n} \, \binom{n+1}{q+1}\sqrt{\var \big[\hat{X}_{q+2}\big]}\Bigg]^2 \notag \\
& \qquad\qquad \le 4 \Bigg[(n+1)\max_{0\le q \le n} \, \binom{n+1}{q+1}\sqrt{\var \big[\hat{X}_{q+2}\big]}\Bigg]^2 \le 4 (n+1)^2 \max_{0\le q \le n} \, \binom{n+1}{q+1}^{\!2} \var \big[\hat{X}_{q+2}\big].
\end{align}
Using now Eq.~\eqref{var-xq}, we obtain
\begin{align}
\var[\hat{F}_n] & \le 4 (n+1)^2 \max_{0\le q \le n} \, \binom{n+1}{q+1}^{\!\!2} \, \sum_{k=1}^{q+2}
    \frac{(q+2)!^2 2^{kN}}{k!(q+2-k)!^2(M-k+1)^k}
     \tr \big( [O_k^{(q+2)}]^2\big), \label{varfn}
\end{align}
where $O_k^{(q+2)}$ is the $k$-copy operator defined in Eq.~\eqref{oqk}, from the $(q+2)$-copy operator $O^{(q+2)}$ defined in Eq.~\eqref{eq:Oq2}.

Using Chebyshev’s inequality, we can now ensure in particular that $\mathrm{Pr}[|\hat{F}_n - F_n| \geq \epsilon] \leq \delta$, for some given accuracy $\epsilon$ and confidence level $\delta$, if for all $q$, each term in the sum (including the prefactors) is upper-bounded by $\epsilon^2\delta/(q+2)$, or equivalently, if 
\begin{align}
M \ge \max_{1\le k\le q+2\le n+2} \left\{\Bigg( 4 (n+1)^2 \binom{n+1}{q+1}^{\!\!2} \frac{(q+2)\,(q+2)!^2}{k!(q+2-k)!^2}
     \frac{\tr \big( [O_k^{(q+2)}]^2\big)}{\epsilon^2\delta}\Bigg)^{1/k} 2^N + k-1 \right\}. \label{conffn}
\end{align}
The measurement budget $M$ can now be evaluated by estimating $\tr \big( [O_k^{(q+2)}]^2\big)$ (either analytically or numerically) which is expressed as function of the state $\rho$ and the operator $A$. Analogous to the expression of Eq.~\eqref{conf-xq}, we could potentially expect for the above confidence interval bound, $n+2$ regimes where the measurements scale as $M \propto \alpha 2^N/\epsilon^{\frac{2}{k}}$ with $k = 1,\dots, n+2$.
\section{Explicit error bounds and required number of measurements for $F_0$ and $F_1$}
\label{app_f0_f1}
In this section, we apply the recipe provided in Appendix~\ref{app-bound-fn} to evaluate the number of measurements required to estimate $F_0$ and $F_1$ for a given value of  $\epsilon$, and $\delta$.
We write down $F_0$ and $F_1$ following Eq.~\eqref{fn-gen} on two copies and three copies of the state $\rho$ as
\begin{equation}
    F_0 = 2 X_2 = 2\,\tr \big(O^{(2)} \rho^{\otimes 2} \big) \qquad \text{and} \qquad F_1 = 4\,X_2 -2 X_3 = 4\,\tr \big(O^{(2)} \rho^{\otimes 2} \big) - 2\,\tr \big(O^{(3)} \rho^{\otimes 3} \big).
\end{equation}
with
\begin{eqnarray}
    O^{(2)} &=& 2\big[\mathbb{1}\otimes A^2 - A\otimes A\big]\Pi_{(2)}, \label{O20}\\
    O^{(3)} &=& \big[ 2 (\mathbb{1}\otimes \mathbb{1}\otimes A^2) - \mathbb{1}\otimes A\otimes A - A\otimes \mathbb{1}\otimes A \big]\Pi_{(3)}. \label{O31}
\end{eqnarray}
We start by bounding the variance of $F_0$ which can be easily achieved by using Eq.~\eqref{varfn}. Here we would like to calculate the necessary two-copy operators $O^{(2)}_k$ for different orders of $k = 1,2$ with $O^{(2)}$ defined in Eq.~\eqref{O20}, in order to finally obtain an error bound on $F_0$ using Eq.~\eqref{conffn}. As remarked earlier, the permutation operators $\pi$ on two-copies are the identity and the swap. We can thus write the scrambled operator $\bar{O}^{(2)}$ as:
\begin{equation}
    \bar{O}^{(2)} = \frac{1}{2!} \sum_{\pi} \pi^\dag O^{(2)} \pi = [\mathbb{1}\otimes A^2 - A\otimes A]\Pi_{(2)} + \Pi_{(2)}[\mathbb{1}\otimes A^2 - A\otimes A] = (A\otimes\mathbb{1} - \mathbb{1}\otimes A)^2 \, \Pi_{(2)}.
\end{equation}

This allows us to write the operators $O^{(2)}_k$ as:
\begin{eqnarray}
    O^{(2)}_1 = \mathrm{Tr}_{\{2\}} \big( \bar{O}^{(2)} [\mathbb{1} \otimes \rho] \big) = \rho A^2 + A^2 \rho - 2 A\rho A \, , \qquad O^{(2)}_2 =  \bar{O}^{(2)},
\end{eqnarray}
and using the cyclic nature of the trace we can evaluate the following terms
\begin{eqnarray}
    \tr([O^{(2)}_1]^2) &=& \tr(2 \rho^2 A^4 + 6 \rho A^2 \rho A^2 - 8 \rho A^3 \rho A) \nonumber \\
    \tr([O^{(2)}_2]^2) &=& 2\tr(A^4)2^N + 6\tr(A^2)^2  - 8 \tr(A)\tr(A^3) \label{eq:Tr_O21squared_Tr_O22squared}
\end{eqnarray}
With the above concrete expressions and Eq.~\eqref{varfn}, we can bound the variance of $\hat{F}_0$ as
\begin{equation}
    \var[\hat{F}_0] \le 16\frac{2^N}{M} \mathrm{Tr}([O^{(2)}_1]^2) + 8\frac{2^{2N}}{(M-1)^2}\mathrm{Tr([O^{(2)}_2]^2)}. \label{varf0}
\end{equation}

One can see from Eq.~\eqref{varf0} that there exists two error scalings coming from the linear $\tr([O^{(2)}_1]^2)$ and quadratic terms $\tr([O^{(2)}_2]^2)$. 
With help of Eq.~\eqref{conffn}, we can assert $F_0$ with a confidence level of at least $1-\delta$ and $|\hat{F}_0 - F_0| \leq \epsilon$ for any $\delta,\,\epsilon > 0$, by satisfying
\begin{equation}
    M \ge \mathrm{max}\Bigg\{ 32\frac{\mathrm{Tr}([O^{(2)}_1]^2)}{\epsilon^2\delta} 2^N , 4\frac{\sqrt{\mathrm{Tr([O^{(2)}_2]^2)}}}{\epsilon\sqrt{\delta}}2^N + 1 \Bigg \}. \label{conff0}
\end{equation}

\medskip

Similarly, to bound the variance of $\hat{F}_1$, we need to additionally calculate the third order terms $O^{(3)}_k$ with $k = 1,\,2,\,3$ for the three copy operator $O^{(3)}$ to obtain the global error bound on $\hat{F}_1$. Below, we would like to focus mainly on the explicit expression of $\tr([O^{(3)}_1]^2)$ which could be the relevant dominant contribution in the high accuracy regime $\epsilon \to 0$. 

The average over of the 6 three-copy permutation operators $\pi$ for the scrambled operator $\bar{O}^{(3)}$ can be written as $\frac{1}{3!}\sum_{\pi} \pi^\dag O^{(3)} \pi = \frac{1}{3!}\sum_{\pi} (\pi^\dag \big[ 2 (\mathbb{1}\otimes \mathbb{1}\otimes A^2) - \mathbb{1}\otimes A\otimes A - A\otimes \mathbb{1}\otimes A \big] \pi )( \pi^\dag \Pi_{(3)} \pi)$). Noting that $\pi^\dag \Pi_{(3)} \pi$ is either equal to $\Pi_{(3)}$ or to $\Pi_{(3)}^2$ for all $\pi$, we find
\begin{equation}
    \bar{O}^{(3)} = \frac{1}{3}\Big( \mathbb{1}\otimes\mathbb{1}\otimes A^2 + \mathbb{1}\otimes A^2\otimes\mathbb{1} + A^2\otimes\mathbb{1}\otimes\mathbb{1} - \mathbb{1}\otimes A \otimes A - A \otimes \mathbb{1}\otimes A - A \otimes A\otimes \mathbb{1}\Big) \Big( \Pi_{(3)} + \Pi_{(3)}^2\Big)
\end{equation}
and we can then calculate the terms $O^{(3)}_k$ as
\begin{eqnarray}
    O^{(3)}_1 &=& \tr_{\{2,3\}}\big(\bar{O}^{(3)}[\mathbb{1} \otimes \rho^{\otimes 2}]\big) = \frac{2}{3}\Big( \rho A^2\rho + \rho^2 A^2 + A^2\rho^2 - A\rho^2 A - \rho A\rho A - A\rho A\rho\Big) \nonumber \\[1mm]
    O^{(3)}_2 &=& \tr_{\{3\}}\big(\bar{O}^{(3)}[\mathbb{1}^{\otimes 2} \otimes \rho]\big) = \frac{1}{3}\Big( \rho A^2\otimes\mathbb{1} + \mathbb{1}\otimes\rho A^2 + A^2\rho \otimes\mathbb{1} + \mathbb{1}\otimes A^2\rho + A^2 \otimes\rho + \rho\otimes A^2 \nonumber \\[-1mm]
    && \hspace{40mm} - A\rho A\otimes\mathbb{1} - \mathbb{1}\otimes A\rho A - A\rho \otimes A - A\otimes  A\rho - \rho A\otimes A - A\otimes \rho A\Big)\Pi_{(2)} \nonumber \\[1mm]
    O^{(3)}_3 &=& \bar{O}^{(3)}. \label{eq:O31_O32_O33}
\end{eqnarray}

Using Eq.~\eqref{varfn}, we can express the upper bound on $\var[\hat{F}_1]$ as
\footnotesize
\begin{align}
\var[\hat{F}_1] 
     & \le \max \bigg\{ 256 \frac{2^{N}}{M} \tr \big( [O_1^{(2)}]^2\big) + 128 \frac{2^{2N}}{(M{-}1)^2} \tr \big( [O_2^{(2)}]^2\big) ,144 \frac{2^{N}}{M}\tr \big( [O_1^{(3)}]^2\big)
     + 288 \frac{2^{2N}}{(M{-}1)^2}\tr \big( [O_2^{(3)}]^2\big)
     + 96 \frac{2^{3N}}{(M{-}2)^3}\tr \big( [O_3^{(3)}]^2\big) \bigg\}.
\end{align}
\normalsize
Following the Chebyshev’s inequality and from Eq.~\eqref{conffn} we can provide the necessary error bounds where we can assert $F_1$ with a confidence level of at least $1-\delta$ and $|\hat{F}_1 - F_1| \leq \epsilon$ for any $\delta,\,\epsilon > 0$, when we take the number of measurements
\footnotesize
\begin{equation}
M \ge \max \Bigg\{ \frac{\max \big\{512\, \tr \big( [O_1^{(2)}]^2\big), 432\, \tr \big( [O_1^{(3)}]^2\big)\big\}}{\epsilon^2\delta} 2^N,
     \frac{\max \big\{16\, \sqrt{\tr \big( [O_2^{(2)}]^2\big)}, 12\sqrt{6}\, \sqrt{\tr \big( [O_2^{(3)}]^2\big)} \big\}}{\epsilon\sqrt{\delta}} 2^N {+} 1, 
     2\, \frac{6^{\frac23} \,\tr \big( [O_3^{(3)}]^2\big)^{\frac13}}{\epsilon^{\frac23}\delta^{\frac13}} 2^N {+} 2\Bigg\}.\label{conff1}
\end{equation}
\normalsize


For illustration, let us evaluate the expressions of Eq.~\eqref{conff0} and Eq.~\eqref{conff1} in the high accuracy regime $\epsilon\to 0$ (i.e., $k=1)$, and in the context of 
pure $N-$qubit states $\rho = \ket{\mathrm{GHZ}_N}\bra{\mathrm{GHZ}_N}$ and the operator $A = \frac{1}{2} \sum_{l = 1}^{N} \sigma_z^{(l)}$.
In this case, we find%
\footnote{This can be seen by first noting that $A\ket{\mathrm{GHZ}_N^\pm} = \frac{N}{2}\ket{\mathrm{GHZ}_N^\mp}$, with $\ket{\mathrm{GHZ}_N^\pm} = (\ket{0}^{\otimes N}\pm\ket{1}^{\otimes N})/\sqrt{2}$, so that $\bra{\mathrm{GHZ}_N}A^2\ket{\mathrm{GHZ}_N} = N^2/4$, $\bra{\mathrm{GHZ}_N}A^4\ket{\mathrm{GHZ}_N} = N^4/16$, $\bra{\mathrm{GHZ}_N}A\ket{\mathrm{GHZ}_N} = \bra{\mathrm{GHZ}_N}A^3\ket{\mathrm{GHZ}_N} = 0$, and then using Eqs.~\eqref{eq:Tr_O21squared_Tr_O22squared} and~\eqref{eq:O31_O32_O33}.}
that $\tr \big( [O_1^{(2)}]^2\big) = N^4/2$ and $\tr \big( [O_1^{(3)}]^2\big) = N^4/4$. This shows that both terms are directly proportional to $N^4$, and they do not contain an hidden exponential scaling.
Explicitly we can write Eqs.~\eqref{conff0} and~\eqref{conff1} as $M \ge 16\frac{N^4}{\epsilon^2\delta} 2^N$ and $M \ge 256\frac{N^4}{\epsilon^2\delta} 2^N$, respectively.
Importantly, while $F_1$ requires more measurements than  $F_0$,  the measurement budget $M$ scales $\propto 2^N$ for both quantities (for a fixed `relative' accuracy $\epsilon=\gamma F_Q=\gamma N^2$, $\gamma\to 0$).

\section{Additional numerical simulations}
\label{app_numerics}
\begin{figure}[h]
\begin{minipage}[b]{0.33\linewidth}
\centering
\includegraphics[width=\textwidth]{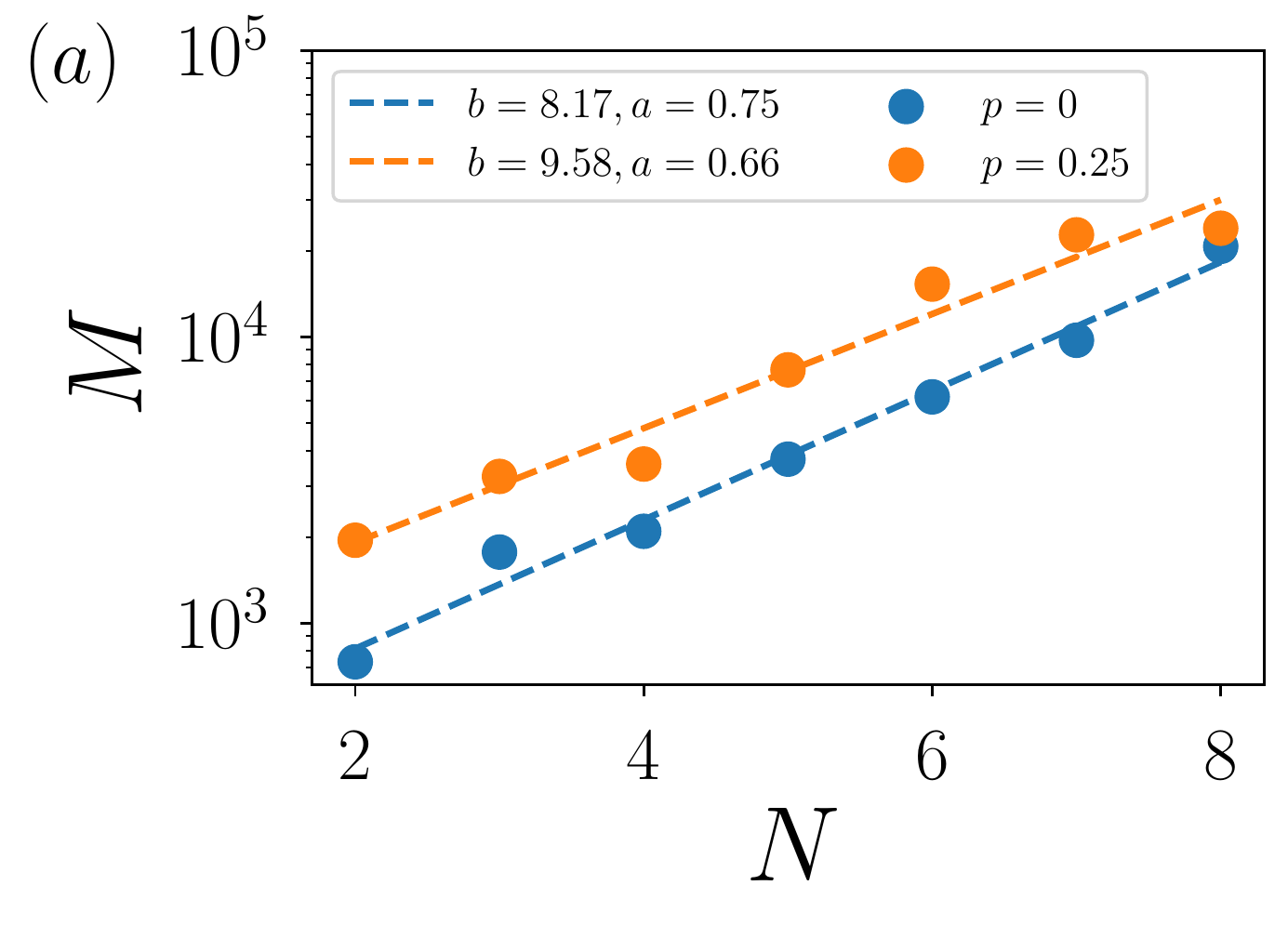}
\end{minipage}
\hskip -1ex
\begin{minipage}[b]{0.33\linewidth}
\centering
\includegraphics[width=\textwidth]{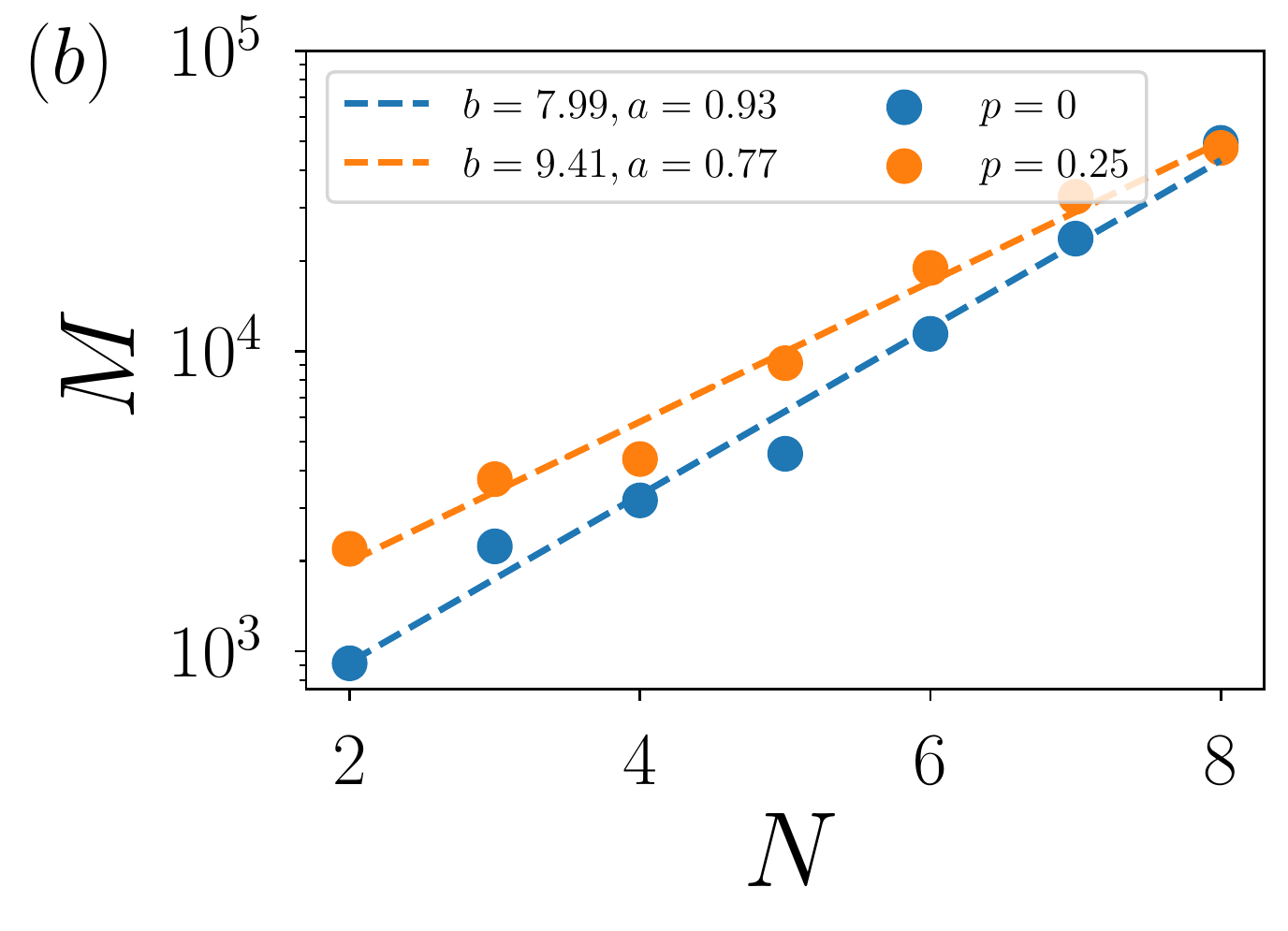}
\end{minipage}
\caption{{\it Scaling of the required number of measurements ---}  Panels (a) and (b) provide the number of measurements $M$ required to estimate $F_0$ and $F_1$ respectively below an error of 0.1 for different values of $p$ of a noisy GHZ state, with respect to $A = \frac{1}{2} \sum_{l = 1}^{N} \sigma_z^{(l)}$. The dashed lines are exponential fits of the type $2^{b +aN}$ highlighting the scaling as a function of the system size $N$. \label{fig:scaling-SM}}
\end{figure}

\noindent
Here, we provide additional numerical simulations of the scalings of the statistical errors on our lower bounds $F_0$ and $F_1$. We consider an $N$-qubit noisy GHZ state $\rho(p) = (1-p)\ket{\mathrm{GHZ}_N}\bra{\mathrm{GHZ}_N} + p\,\mathbb{1}/2^N$ parametrized by the noise strength $p$. The lower bounds $F_0$ and $F_1$ are computed with respect to the observable $A = \frac{1}{2} \sum_{l = 1}^{N} \sigma_z^{(l)}$. We simulate the protocol by applying $M$ local random untaries $u$ followed by measurements in a fixed basis to obtain estimates $\hat{F}_n$. The average statistical error $\mathcal{E}$ is calculated by averaging the relative error $\hat{\mathcal{E}} = |\hat{F}_n - F_n|/F_n$ with $n = 0, \, 1$ over 50 experimental runs for different values of $M$. We find the maximum value of $M$ for which we obtain $\mathcal{E} \leq 0.1$ for different system sizes $N$ by using a linear interpolation function. This is plotted in Fig.~\ref{fig:scaling-SM}, where the scaling exponents are fitted to the obtained values of $M$ vs $N$. 
We see that the exponential fits agree with the exponents found in the scaling analysis of the main text and remain favorable. 
The exponents $a$ and $b$ also highlight the fact that the measurements $M$ needed to obtain an error $\cal{E}$ is less for $F_0$ as compared to $F_1$.  

We now show below in Fig.~\ref{fig:non-rescaled} the non-rescaled plots of Figs.~(2)(b,c)-(3)(b,c) of the main text for GHZ and N00N states mixed with depolarization noise of strength $p = 0.25$. We estimate $F_0$ with a lower statistical error than $F_1$ for the same number of measurements $M$ at the price of the bound $F_0$ being less tight compared to $F_1$. Fig.~\ref{fig:non-rescaled}(a,b) and Fig.~\ref{fig:non-rescaled}(c,d) explicitly highlight the trade-off of the required number of measurements to estimate a tighter bound ($F_1$ compared to $F_0$) with a certain precision. 
\vspace{2cm}
\begin{figure}[h]
\begin{minipage}[b]{0.24\linewidth}
\centering
\includegraphics[width=\textwidth]{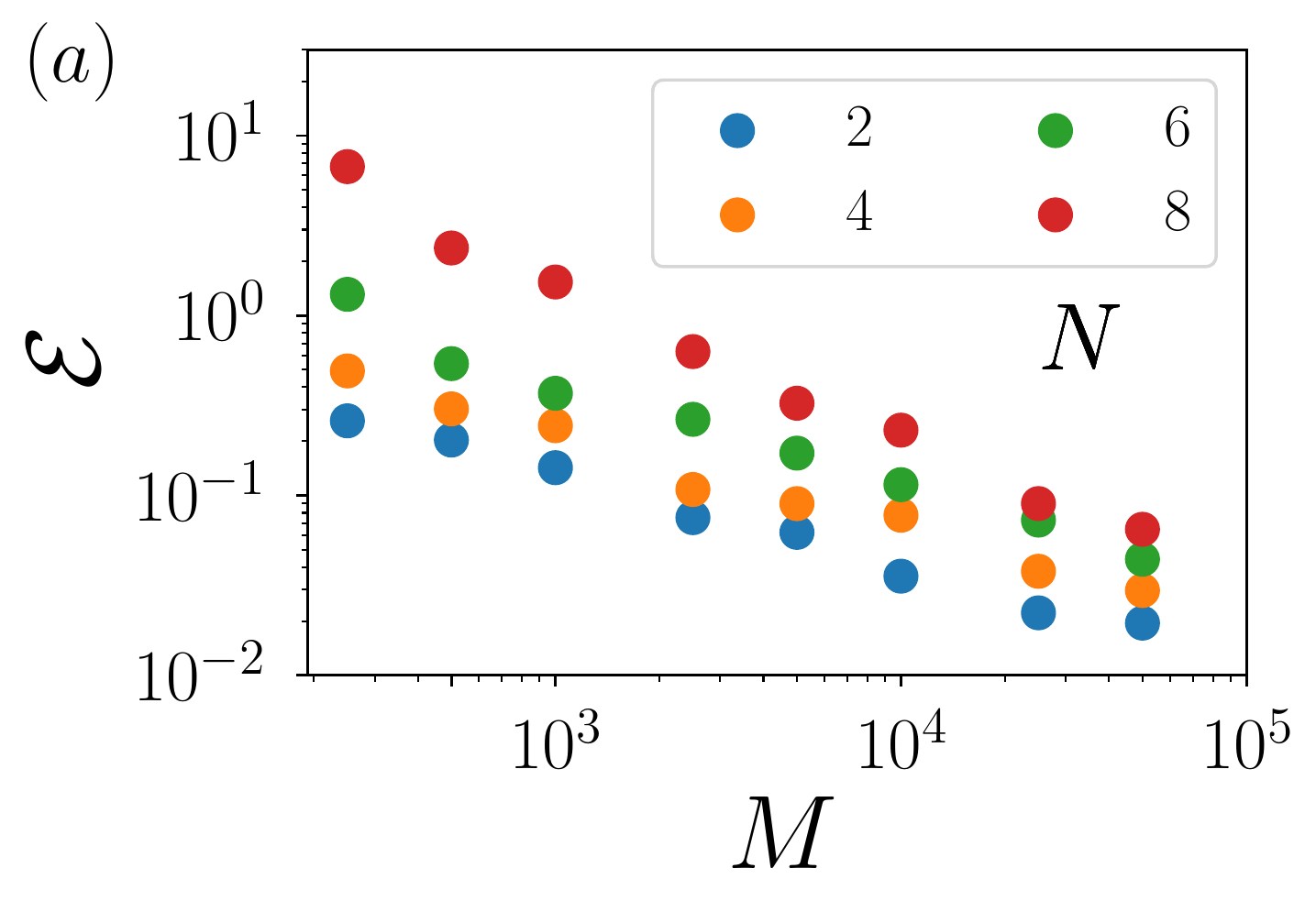}
\end{minipage}
\hskip -1ex
\begin{minipage}[b]{0.24\linewidth}
\centering
\includegraphics[width=\textwidth]{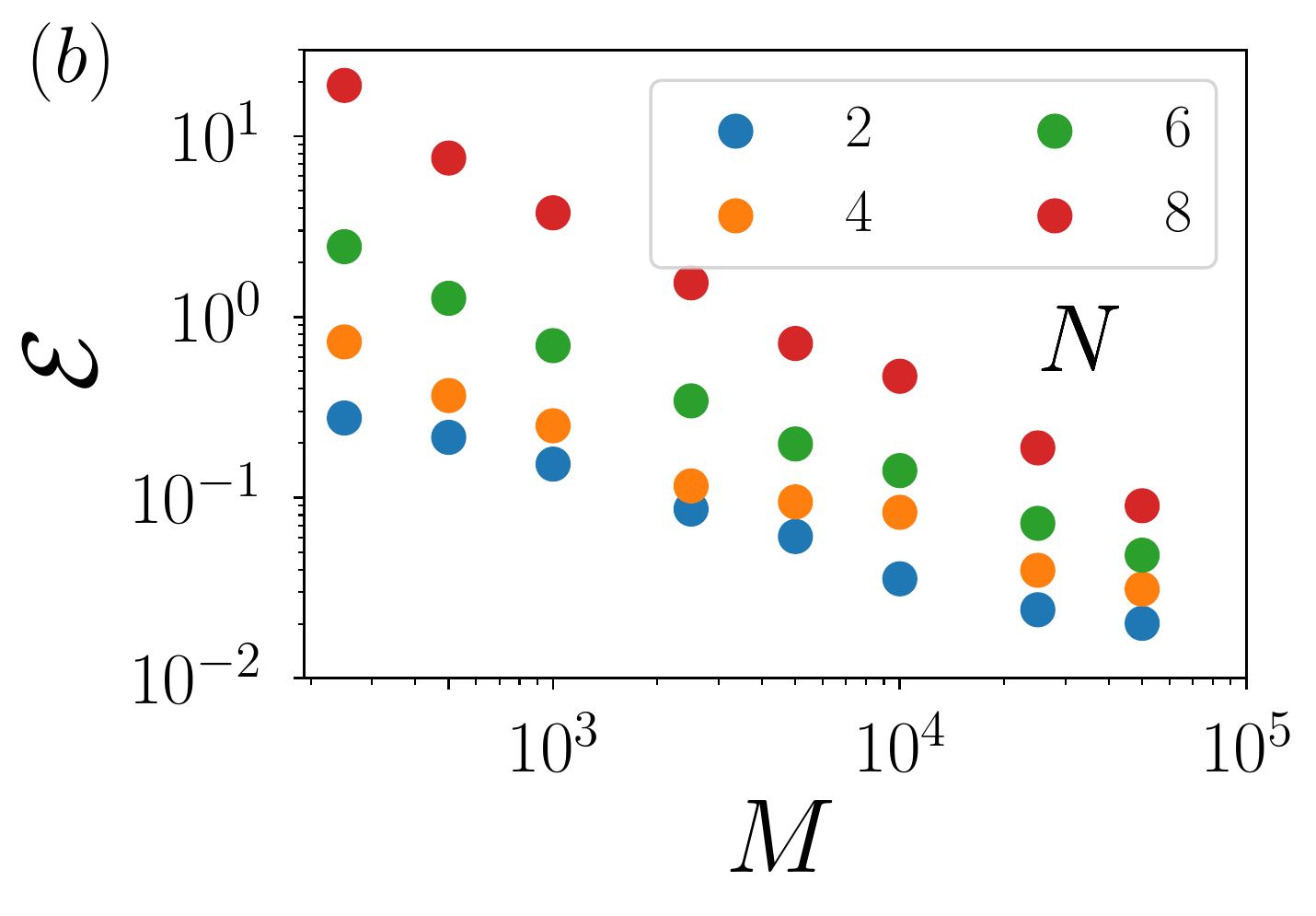}
\end{minipage}
\begin{minipage}[b]{0.24\linewidth}
\centering
\includegraphics[width=\textwidth]{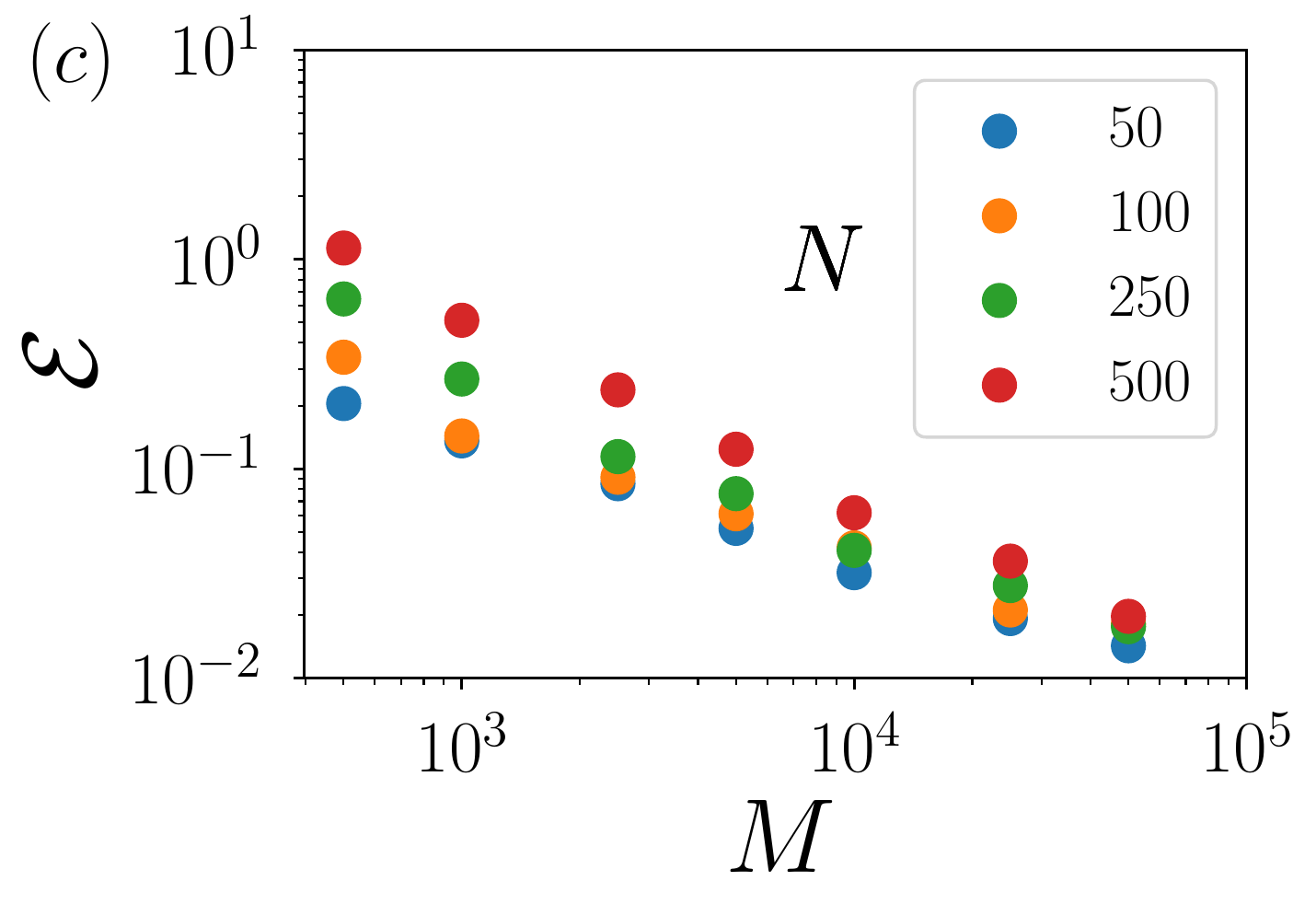}
\end{minipage}
\hskip -1ex
\begin{minipage}[b]{0.24\linewidth}
\centering
\includegraphics[width=\textwidth]{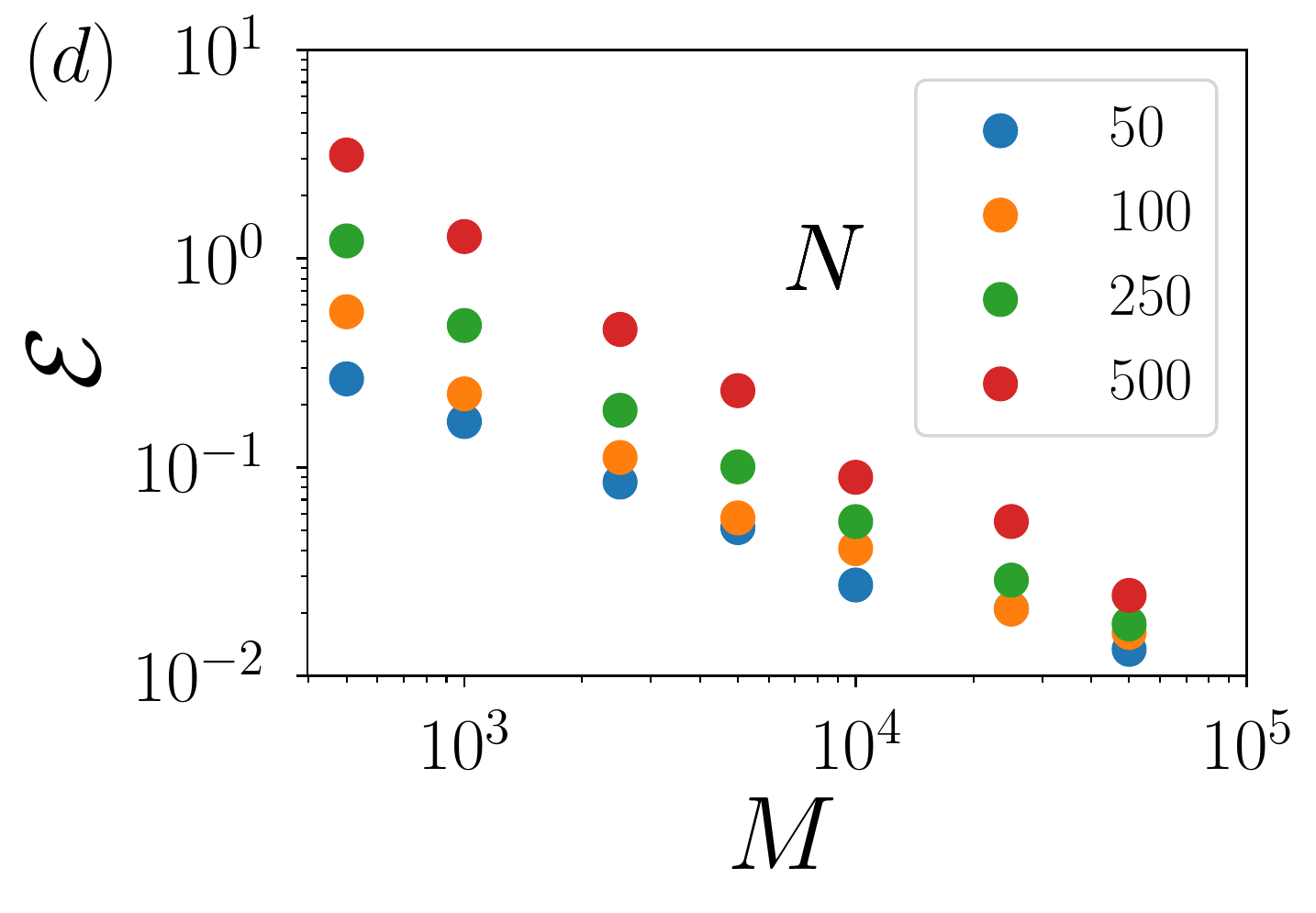}
\end{minipage}
\caption{{\it Statistical error comparisons between $F_0$ and $F_1$ ---}  Panels (a) and (b) GHZ states and (c) and (d) N00N states, show for different system sizes $N$ (see legends), the average statistical error for $F_0$ (panels (a) and (c)) and $F_1$ (panels (b) and (d)) as a function of the number of measurements $M$. \label{fig:non-rescaled}}
\end{figure}

\end{document}